\documentstyle[fleqn,twoside,epsfig]{article} 
\begin{document}
\pagestyle{myheadings}
\markboth{\rm Koonin, Dean, \& Langanke}{\rm Shell Model Monte Carlo}

\noindent
{\huge Results from Shell Model Monte Carlo Studies} \bigskip

\noindent
{\large {\it S. E. Koonin}}

\medskip
\noindent
Kellogg Radiation Laboratory, 106-38\\ California Institute of Technology,
Pasadena, CA 91125 USA

\medskip
\noindent
{\large {\it D. J. Dean }}

\medskip
\noindent
Physics Division, Oak Ridge National Laboratory, \\
Oak Ridge, TN 37831-6373 USA

\medskip
\noindent
{\large {\it K. Langanke}}

\medskip
\noindent
Institute for Physics and Astronomy, Aarhus University, Denmark

\bigskip
\noindent {\sc key words}:\quad Nuclei, Shell Model, Monte Carlo

\bigskip
\hrule
\bigskip
\begin{abstract}
We review results obtained using Shell Model Monte Carlo (SMMC) techniques.
These methods reduce the imaginary-time many-body evolution operator to a
coherent superposition of one-body evolutions in fluctuating one-body fields;
the resultant path integral is evaluated stochastically. After a brief review
of the methods, we discuss a variety of nuclear physics applications. These
include studies of the ground-state properties of $pf$-shell nuclei,
Gamow-Teller strength distributions, thermal and rotational pairing
properties of nuclei near $N=Z$, $\gamma$-soft nuclei, and $\beta\beta$-decay
in ${}^{76}$Ge. Several other illustrative calculations are also reviewed.
Finally, we discuss prospects for further progress in SMMC and related
calculations.
\end{abstract}

\bigskip
\hrule
\bigskip
\tableofcontents

\bigskip

\section{Introduction}

The notion of independent particles moving in a common one-body potential is
central to our description of atoms, metals, and hadrons. It is also realized
in nuclei, and the shell structure associated with the magic numbers was
first put on a firm basis in 1949, when the magic numbers were explained by
an harmonic oscillator spectrum with a strong, inverted (with respect to the
atomic case) spin-orbit potential \cite{Haxel}.

But nuclei differ from the other quantal systems cited above in that the
residual interaction between the valence fermions is strong and so severely
perturbs the naive single-particle picture. This interaction mixes together
many different configurations to produce the true eigenstates and, because of
its coherence, there emerge phenomena such as pairing, modification of sum
rules, deformation, and collective rotations and vibrations. An accurate
treatment of the residual interaction is therefore essential to properly
describe nuclei.

A hierachy of models has been developed to treat various aspects of the
mean field and correlations; Skyrme Hartree-Fock\cite{HF} for
the mean field, Hartree-Fock Bogolioubov for the pairing correlations
\cite{HFB}, and RPA \cite{RPA} for particle-hole correlations.
But short of a complete solution of the many-body problem, the shell
model is regarded as the most fundamental approach for studying nuclear
structure, provided that the single-particle space in which calculations are
performed is large enough. Success was demonstrated early in the $0p$-shell
\cite{Kurath}, in the $0s$-$1d$ \cite{Wild84}, in the $0f_{7/2}$ region
\cite{McGrory}, and more recently in the $1p0f$-shell using interactions
developed by Kuo and Brown \cite{Kuo} and modified by Zuker and Poves
\cite{Zuker}.

The traditional numerical approach to the nuclear shell model is to
diagonalize the hamiltonian matrix within a limited many-body basis 
\cite{Livermore}. Direct diagonalization gives the wave functions which can
then be used to calculate the properties of specific levels. The shell model
treated in this way has been quite successful in describing $sd$-shell nuclei
\cite{Wild88}, and has also been used to successfully describe various
properties of the lower $pf$-shell \cite{Caurier94,Poves}. However, the
direct-diagonalization procedure is limited by memory requirements that grow
combinatorially with the size of the single-particle space, $N_s$, and with
the number of valance particles, $N_v$. In the $pf$-shell where ${}^{40}$Ca
is the core, the largest systematic study thus far treats $A=48,49$
\cite{Caurier94}, an achievement that required approximately five generations
of technological and software improvement.

To circumvent the limitations of direct diagonalization, we have proposed an
alternative treatment of the shell model
\cite{Johnson,Lang,Ormand,Alhassid,Koonin96}. Our methods are based on a
path-integral formulation of the imaginary-time many-body propagator,
$\exp(-\beta H)$, where $H$ is the shell model hamiltonian, and $\beta$ is
the reciprocal of the temperature. We employ the Hubbard-Stratonovich (HS)
transformation \cite{HST} to recast the two-body terms in the exponential
into one-body terms with an integration over auxiliary fields. Thus the
nucleons are seen as independent particles moving about in fluctuating fields
\cite{Sugiyama}. Such auxiliary field methods have also been applied to
condensed matter systems such as the Hubbard model \cite{Linden,Hirsch},
yielding important information about electron correlations and magnetic
properties.

Having circumvented the diagonalization of the hamiltonian, we are left with
a large multidimensional integration to perform, for which we use the Monte
Carlo sampling techniques of Metropolis {\it et~al} \cite{Metropolis}.
Realistic nuclear hamiltonians often have a sign problem that hampers the
Monte Carlo quadrature. We have overcome this by an extrapolation procedure
from a family of hamiltonians, all of which have no sign problem and are
close to the full hamiltonian. It is important to note that these
calculations are carried out on large scale parallel computers, a
technology that is essential to implement these methods.

In this article, we briefly outline SMMC methods in Section~II, then
concentrate our discussion on the physics results obtained. We discuss in
Section~III our major results, which include ground state properties of
$pf$-shell nuclei, Gamow-Teller strengths and distributions, thermal and
rotational pairing properties of nuclei near $N=Z$, $\gamma$-soft nuclei, and
$\beta\beta$-decay. In Section~IV, several other illustrative studies are
reviewed. We conclude in Section~V with a discussion of future prospects. Our
coverage of the field extends through February 1997.

\section{Methods}

The nuclear shell model is defined by a set of spin-orbit coupled
single-particle states with quantum numbers $ljm$ denoting the orbital
angular momentum ($l$) and the total angular momenta ($j$) and its
$z$-component, $m$. Although non-spherical one-body potentials are a common
efficiency used in describing deformed nuclei, for the rotationally invariant
hamiltonians used in SMMC so far, these states have energies
$\varepsilon_{lj}$ that are independent of $m$. The single-particle states
and energies may be different for neutrons and protons, in which case it is
convenient to include also the isospin component $t_3=\pm 1/2$ in the state
description. We will use the label $\alpha$ for the set of quantum numbers
$ljm$ or $ljmt_3$, as appropriate.

In the following we briefly outline the formalism of the SMMC method. We
begin with a brief description of statistical mechanics techniques used in
our approach, then 
discuss the Hubbard-Stratonovich transformation, and end with a
discussion of Monte Carlo sampling procedures. We refer the reader to
previous works \cite{Lang,Koonin96} for a more detailed exposition.

\subsection{Observables}

SMMC methods rely on an ability to calculate the imaginary-time many-body
evolution operator, $\exp (-\beta H)$, where $\beta$ is a real
$c$-number. The many-body hamiltonian can be written schematically as
\begin{equation}
 H=\varepsilon {\cal O}
+{1\over2}V {\cal O} {\cal O}\;,
\label{eq_a}
\end{equation}
where ${\cal O}$ is a density operator, $V$ is
the strength of the two-body interaction, and $\varepsilon$ a single-particle
energy. In the full problem, there are many such quantities with various
orbital indices that are summed over, but we omit them here for the sake of
clarity.

While the SMMC technique does not result in a complete solution to the
many-body problem in the sense of giving all eigenvalues and eigenstates of
$ H$, it can result in much useful information. For example, the
expectation value of some observable $ \Omega$ can be obtained by
calculating
\begin{equation}
\langle  \Omega\rangle= {{\rm Tr}\,
e^{-\beta H} \Omega\over {\rm Tr}\,
e^{-\beta H}}\;.
\label{eq_b}
\end{equation}
Here, $\beta\equiv T^{-1}$ is interpreted as the inverse of the temperature
$T$, and the many-body trace is defined as
\begin{equation}
{\rm Tr}\, X\equiv\sum_i \langle i\vert  X\vert i\rangle\;,
\label{eq_c}
\end{equation}
where the sum is over many-body states of the system. In the canonical
ensemble, this sum is over all states with a specified number of nucleons
(implemented by ``number projection'' \cite{Lang,Koonin96}), while the grand
canonical ensemble introduces a chemical potential and sums over {\it all}
many-body states.

In the limit of low temperature ($T\rightarrow0$ or
$\beta\rightarrow\infty$), the canonical trace reduces to a ground state
expectation value. Alternatively, if $\vert \Phi\rangle$ is a many-body trial
state not orthogonal to the exact ground state, $\vert\Psi\rangle$, then
$e^{-\beta H}$ can be used as a filter to refine $\vert\Phi\rangle$ to
$\vert\Psi\rangle$ as $\beta$ becomes large. An observable can be calculated
in this ``zero temperature'' method as
\begin{equation}
{\langle\Phi\vert e^{-{\beta\over2} H} \Omega
e^{-{\beta\over2} H} \vert\Phi\rangle\over
\langle\Phi\vert e^{-\beta H}\vert\Phi\rangle}
{}~\hbox{$\longrightarrow\hskip-20pt{}^{\beta\rightarrow\infty}$}
{\langle\Psi\vert \Omega\vert\Psi\rangle\over
\langle\Psi\vert\Psi\rangle}\;.
\label{eq_d}
\end{equation}
If $ \Omega$ is the hamiltonian, then (\ref{eq_d}) at $\beta=0$ is the
variational estimate of the energy, and improves as $\beta$ increases. Of
course, the efficiency of the refinement for any observable depends upon the
degree to which $\vert\Phi\rangle$ approximates $\vert\Psi\rangle$.

Beyond such static properties, $e^{-\beta H}$ allows us to obtain some
information about the dynamical response of the system. For an operator $
\Omega$, the response function $R_\Omega(\tau)$ in the canonical ensemble is
defined as
\begin{equation}
R_\Omega(\tau)\equiv {{\rm Tr}\,
e^{-(\beta-\tau) H}  \Omega^\dagger
e^{-\tau H} \Omega\over {\rm Tr}\,
e^{-\beta H}}\equiv \langle\Omega^\dagger(\tau)\Omega(0)\rangle,
\label{eq_e}
\end{equation}
where $\Omega^\dagger(\tau)\equiv e^{\tau H} \Omega^\dagger
e^{-\tau H}$ is the imaginary-time Heisenberg operator. Interesting
choices for $\Omega$ are the annihiliation operators 
for particular orbitals, the
Gamow-Teller, $M1$, or quadrupole moment, etc. Inserting complete sets of
$A$-body eigenstates of $ H$ ($\{\vert i\rangle,\vert f\rangle\}$ with
energies $E_{i,f}$) shows that
\begin{equation}
R_\Omega(\tau) ={1\over Z}\sum_{if}
e^{-\beta E_i} \vert\langle f\vert\Omega\vert i\rangle\vert^2 e^{-\tau(E_f-E_i)},
\label{eq_f}
\end{equation}
where $Z=\sum_i e^{-\beta E_i}$ is the partition function. Thus,
$R_\Omega(\tau)$ is the Laplace transform of the strength function
$S_\Omega(E)$:
\begin{eqnarray}
R_\Omega(\tau)& = & \int^\infty_{-\infty} e^{-\tau E} S_\Omega(E)dE\;; \\
S_\Omega(E)&=& {1\over Z}\sum_{fi} e^{-\beta E_i} \vert\langle f\vert 
\Omega\vert i\rangle\vert^2 \delta(E-E_f+E_i)\;.
\label{eq_g}
\end{eqnarray}

\noindent
Hence, if we can calculate $R_\Omega(\tau)$, $S_\Omega(E)$ can be determined.
Short of a full inversion of the Laplace transform (which is often
numerically difficult), the behavior of $R_\Omega(\tau)$ for small $\tau$
gives information about the energy-weighted moments of $S_\Omega$. In
particular,
\begin{equation}
R_\Omega(0)=\int^\infty_{-\infty} S_\Omega (E) dE=
{1\over Z}\sum_i e^{-\beta E_i} \vert\langle f\vert {\Omega}\vert
i\rangle\vert^2=
\langle {\Omega}^\dagger{\Omega}\rangle_A
\label{eq_h}
\end{equation}
is the total strength,
\begin{equation} -R_\Omega^\prime (0)=
\int^\infty_{-\infty} S_\Omega (E) EdE=
{1\over Z}\sum_{if} e^{-\beta E_i} \vert\langle f\vert \Omega\vert
i\rangle\vert^2 (E_f-E_i)
\label{eq_i}
\end{equation}
is the first moment (the prime denotes differentiation with respect to
$\tau$).

It is important to note that we usually cannot obtain detailed spectroscopic
information from SMMC calculations. Rather, we can calculate expectation
values of operators in the thermodynamic ensembles or the ground state.
Occasionally, these can indirectly furnish properties of excited states. For
example, if there is a collective $2^+$ state absorbing most of the $E2$
strength, then the centroid of the quadrupole response function will be a
good estimate of its energy. But, in general, we are without the numerous
specific excitation energies and wavefunctions that characterize a direct
diagonalization. This is both a blessing and a curse. The former is that for
the very large model spaces of interest, there is no way in which we can deal
explicitly with all of the wavefunctions and excitation energies. Indeed, we
often don't need to, as experiments only measure average nuclear properties
at a given excitation energy. The curse is that comparison with detailed
properties of specific levels is difficult. In this sense, the SMMC method is
complementary to direct diagonalization for modest model spaces, but is the
only method for treating very large problems.

\subsection{The Hubbard-Stratonovich transformation}

It remains to describe the Hubbard-Stratanovich ``trick'' by which
$e^{-\beta H}$ is managed. In broad terms, the difficult many-body
evolution is replaced by a superposition of an infinity of tractable one-body
evolutions, each in a different external field, $\sigma$. Integration over
the external fields then reduces the many-body problem to quadrature.

To illustrate the approach, let us assume that only one operator ${\cal
O}$ appears in the hamiltonian (\ref{eq_a}).  Then all of the difficulty
arises from the two-body interaction, that term in $ H$ quadratic in
${\cal O}$. If $ H$ were solely linear in ${\cal O}$, we would
have a one-body quantum system, which is readily dealt with. To linearize the
evolution, we employ the Gaussian identity
\begin{equation}
e^{-\beta H}=
\sqrt{\beta \mid V\mid \over 2\pi} \int^\infty_{-\infty} d\sigma
e^{-{1\over2}\beta \mid V\mid \sigma^2}
e^{-\beta h};\;\;\;
 h= \varepsilon {\cal O} +s V\sigma{\cal O}\;.
\label{eq_j}
\end{equation}
Here, $ h$ is a one-body operator associated with a $c$-number field
$\sigma$, and the many-body evolution is obtained by integrating the one-body
evolution $ U_\sigma\equiv e^{-\beta h}$ over all $\sigma$ with a
Gaussian weight. The phase, $s$, is $1$ if $V<0$ or $i$ if $V>0$.
Equation~(\ref{eq_j}) is easily verified by completing the square in the
exponent of the integrand; since we have assumed
that there is only a single operator ${\cal O}$, there is no need to
worry about non-commutation.

For a realistic hamiltonian, there will be many non-commuting density
operators ${\cal O}_\alpha$ present, but we can always reduce
the two-body term to diagonal form. Thus for a general two-body interaction
in a general time-reversal invariant form, we write
\begin{equation}
{H}=\sum_\alpha \left(\epsilon^\ast_\alpha {\bar{\cal O}}_\alpha+
\epsilon_\alpha  {\cal O}_\alpha\right)+
{1\over2}\sum_\alpha V_\alpha \left\{  {\cal O}_\alpha,
 {\bar{\cal O}}_\alpha\right\}\;,
\label{eq_k}
\end{equation}
where $ {\bar{\cal O}}_\alpha$ is the time reverse of $ {\cal
O}_\alpha$. Since, in general, $[ {\cal O}_\alpha, {\cal
O}_\beta]\not=0$, we must split the interval $\beta$ into $N_t$ ``time
slices'' of length $\Delta\beta\equiv\beta/N_t$,
\begin{equation}
e^{-\beta H}= [e^{-\Delta\beta H}]^{N_t}, \label{eq_l}
\end{equation}
and for each time slice $n=1, \ldots, N_t$ perform a linearization similar to
Eq.~\ref{eq_j} using auxiliary fields $\sigma_{\alpha n}$. Note that because
the various $ {\cal O}_\alpha$ need not commute, the representation of
$e^{-\Delta\beta h}$ must be accurate through order $(\Delta\beta)^2$ to
achieve an overall accuracy of order $\Delta\beta$.

We are now able to write expressions for observables as the ratio of two
field integrals. Thus expectations of observables can be written as
\begin{eqnarray}
\langle \Omega\rangle =
{\int{\cal D} \sigma W_\sigma \Omega_\sigma\over
\int{\cal D} \sigma W_\sigma},\qquad\qquad\qquad\qquad \label{eq_m} \\
\noalign{\noindent where\hfill}
W_\sigma = G_\sigma {\rm Tr}\, U_\sigma\;;\qquad G_\sigma=
e^{-{\Delta\beta}\sum_{\alpha n}\mid V_\alpha\mid \mid\sigma_{\alpha
n}\mid^2}\;;\nonumber \\
\Omega_\sigma= {{\rm Tr}\,  U_\sigma  \Omega\over {\rm Tr}\, 
U_\sigma}\;;\qquad
{\cal D} \sigma \equiv \prod^{N_t}_{n=1}\prod_\alpha d\sigma_{\alpha
n}d\sigma_{\alpha n}^{*}
\left(\Delta\beta\vert V_\alpha\vert\over 2\pi\right), \label{eq_n} \\
\noalign{\noindent and\hfill}
 U_\sigma=  U_{N_t}\ldots  U_2 U_1\;;\qquad
 U_n= e^{-\Delta\beta h_n};\nonumber \\
 h_n =\sum_\alpha \left(\varepsilon_\alpha^{*} +
s_\alpha V_\alpha \sigma_{\alpha n}\right){\bar{\cal O}}_\alpha+
\left(\varepsilon_\alpha +
s_\alpha V_\alpha \sigma_{\alpha n}^{*}\right){\cal O}_\alpha\;.
\label{eq_o}
\end{eqnarray}

\noindent
This is, of course, a discrete version of a path integral over $\sigma$.
Because there is a field variable for each operator at each time slice, the
dimension of the integrals ${\cal D} \sigma$ can be very large, often
exceeding $10^5$. The errors in Eq.~\ref{eq_m} are of order $\Delta\beta$, so
that high accuracy requires large $N_t$ and perhaps extrapolation to
$N_t=\infty$ ($\Delta\beta=0$).

Thus, the many-body observable is the weighted average (weight $W_\sigma$) of
the observable $\Omega_\sigma$ calculated in an ensemble involving only the
one-body evolution $ U_\sigma$. Similar expressions involving two
$\sigma$ fields (one each for $e^{-\tau H}$ and $e^{-(\beta-\tau)
H}$) can be written down for the response function (\ref{eq_e}), and all are
readily adapted to the canonical or grand canonical ensembles or to the
zero-temperature case.

An expression of the form (\ref{eq_m}) has a number of attractive features.
First, the problem has been reduced to quadrature---we need only calculate
the ratio of two integrals. Second, all of the quantum mechanics (which
appears in $\Omega_\sigma$) is of the one-body variety, which is simply
handled by the algebra of $N_s\times N_s$ matrices. The price to pay is that
we must treat the one-body problem for all possible $\sigma$ fields.

\subsection{Monte Carlo quadrature and the sign problem}

The manipulations of the previous sections have reduced the shell model to
quadrature. That is, thermodynamic expectation values are given as the ratio
of two multidimensional integrals over the auxiliary fields. The dimension
$D$ of these integrals is of order $N_s^2N_t$, which can exceed $10^5$ for
the problems of interest. Monte Carlo methods are the only practical means of
evaluating such integrals. In this section, we review those aspects of Monte
Carlo quadrature relevant to the task at hand.

We begin by recasting the ratio of integrals in Eq.~(\ref{eq_m}) as
\begin{equation}
\langle \Omega\rangle =
\int d^D\sigma P_\sigma \Omega_\sigma\;, \label{nu_a}
\end{equation}
where
\begin{equation}
P_\sigma = {W_\sigma\over \int d^D\sigma W_\sigma}\;. \label{nu_b}
\end{equation}
Since $\int d^D\sigma P_\sigma=1$ and $P_\sigma\geq0$, we can think of
$P_\sigma$ as a probability density and $\langle \Omega\rangle$ as the
average of $\Omega_\sigma$ weighted by $P_\sigma$. Thus, if $\{\sigma_s, s=1,
\ldots, S\}$ are a set of $S$ field configurations randomly chosen with
probability density $P_\sigma$, we can approximate $\langle
\Omega\rangle$ as
\begin{equation}
\langle \Omega\rangle\approx
{1\over S}\sum^S_{s=1} \Omega_s\;, \label{nu_c}
\end{equation}
where $\Omega_s$ is the value of $\Omega_\sigma$ at the field configuration
$\sigma_s$. Since this estimate of $\langle \Omega\rangle$ depends upon
the randomly chosen field configurations, it too will be a random variable
whose average value is the required integral. To quantify the uncertainty of
this estimate, we consider each of the $\Omega_s$ as a random variable and
invoke the central limit theorem to obtain
\begin{equation}
\sigma^2_{\langle \Omega\rangle} =
{1\over S}\int d^D\sigma P_\sigma (\Omega_\sigma-\langle
\Omega\rangle)^2\approx
{1\over S^2}\sum^S_{s=1} (\Omega_s-\langle \Omega\rangle)^2\;.
\label{nu_d} 
\end{equation}
This variance varies as $S^{-1/2}$.

We employ the Metropolis, Rosenbluth, Rosenbluth, Teller, and Teller
algorithm \cite{Metropolis}, to generate the field configurations ${\cal
O}_s$, which requires only the ability to calculate the weight function for a
given value of the integration variables.  This method requires that the
weight function $W_\sigma$ must be real and non-negative. Unfortunately, many
of the hamiltonians of physical interest suffer from a sign problem, in that
$W_\sigma$ is negative over significant fractions of the integration volume.
To understand the implications of this, let us rewrite Eq.~(\ref{nu_a}) as
\begin{equation}
\langle { \Omega}\rangle =
\int d^D\sigma P_\sigma \Phi_\sigma \Omega_\sigma\;,
\end{equation}
where
$$
P_\sigma=
{{\mid W_\sigma\mid}\over{\int d^D\sigma\mid W_\sigma \mid \Phi_\sigma}} \;,
$$
and $\Phi_\sigma =W_\sigma/\mid W_\sigma\mid $ is the sign of the real part
of $W_\sigma$. (Note that since the partition function is real, we can
neglect the imaginary part.) Since $\mid W_\sigma \mid$ is non-negative by
definition, we can interpret it, suitably normalized, as a probability
density, so that upon rewriting (\ref{nu_a}) as
\begin{equation}
\langle { \Omega} \rangle =
{{\int d\sigma \mid W_\sigma\mid \Phi_\sigma {\Omega}_\sigma}
\over{\int d\sigma \mid W_\sigma \mid \Phi_\sigma}}=
{{\langle \Phi  {\Omega}\rangle}\over{\langle\Phi\rangle}}\;,
\end{equation}
we can think of the observable as a ratio in which the numerator and
denominator can be separately evaluated by MC quadrature. Leaving aside the
issue of correlations between estimates of these two quantities (they can
always be evaluated using separate Metropolis walks), the fractional variance
of $\langle { \Omega}\rangle$ will be
\begin{equation}
{{\sigma_\Omega}\over{\langle { \Omega}\rangle}} =
\sqrt{ {{\langle { \Omega}^2\rangle}\over{\langle\Phi {
\Omega}\rangle^2}}+ 
{{1}\over{\langle\Phi\rangle^2}}-2 \; , }
\end{equation}
which becomes unacceptably large as the average sign $\langle\Phi\rangle$
approaches zero. The average sign of the weight thus determines the
feasibility of naive MC quadrature.  In most cases $\langle\Phi\rangle$
decreases exponentially with $\beta$ or with the number of time slices
\cite{Guber92}.

It has been shown \cite{Lang} that for even-even and $N=Z$ nuclei there is no
sign problem for hamiltonians if all $V_\alpha\leq 0$.  Such forces include
reasonable approximations to the realistic hamiltonian like pairing plus
multipole interactions.  However, for an arbitrary hamiltonian, we are not
guaranteed that all $V_\alpha \leq 0$ (see for example \cite{Alhassid,Lang}).
However, we may expect that a {\it realistic} hamiltonian will be dominated
by terms like those of the schematic pairing plus multipole force (which is,
after all, why the schematic forces were developed) so that it is, in some
sense, close to a hamiltonian for which the MC is directly applicable. Thus,
the ``practical solution'' to the sign problem presented in
Ref.~\cite{Alhassid} is based on an extrapolation of observables calculated
for a ``nearby'' family of hamiltonians whose integrands have a positive
sign. Success depends crucially upon the degree of extrapolation required.
Empirically, one finds that, for all of the many realistic interactions
tested in the $sd$- and $pf$-shells, the extrapolation required is modest,
amounting to a factor-of-two variation in the isovector monopole pairing
strength.

Based on the above observation, it is possible to decompose $ H$ in
Eq.~\ref{eq_k} into its ``good'' and ``bad'' parts, ${ H}= { H}_G+
{ H}_B$, with 
\begin{eqnarray} 
{ H}_G&=&\sum_\alpha(\epsilon_\alpha^\ast 
{\bar {\cal O}}_\alpha+\epsilon_\alpha { {\cal O}}_\alpha)+ 
{1\over2}\sum_{V_\alpha<0}V_\alpha \left\{{ {\cal O}}_\alpha, 
{\bar {\cal O}}_\alpha\right\} \nonumber \\ 
{ H}_B&=&{1\over2}\sum_{V_\alpha>0}V_\alpha 
\left\{ { {\cal O}}_\alpha, {\bar{\cal O}}_\alpha \right\}. 
\end{eqnarray} 
\noindent 
The ``good'' hamiltonian ${ H}_G$ includes, in addition to the one-body
terms, all the two-body interactions with $V_\alpha \leq0$, while the ``bad''
hamiltonian ${ H}_B$ contains all interactions with $V_\alpha>0$. By
construction, calculations with ${ H}_G$ alone have $\Phi_\sigma\equiv1$
and are thus free of the sign problem.

We define a family of hamiltonians ${ H}_g$ that depend on a continuous
real parameter~$g$ as ${ H}_g=f(g){ H}_G+g { H}_B$, so that
${ H}_{g=1}={ H}$, and $f(g)$ is a function with $f(1)=1$ and
$f(g<0)>0$ that can be chosen to make the extrapolations less severe. (In
practical applications $f(g)=1-(1-g)/\chi$ with $\chi\approx4$ has been found
to be a good choice.) If the $V_\alpha$ that are large in magnitude are
``good,'' we expect that ${ H}_{g=0}={ H}_G$ is a reasonable starting
point for the calculation of an observable $\langle{{ \Omega} }\rangle$.
One might then hope to calculate $\langle{{ \Omega}}\rangle_g={\rm
Tr}\,({ \Omega}e^{-\beta  H_g})/{\rm Tr}\,(e^{-\beta  H_g})$ for
small $g>0$ and then to extrapolate to $g=1$, but typically
$\langle\Phi\rangle$ collapses even for small positive $g$. However, it is
evident from our construction that ${ H}_g$ is characterized by
$\Phi_\sigma\equiv1$ for any $g\leq 0$, since all the ``bad'' $V_\alpha(>0)$
are replaced by ``good'' $g V_\alpha<0$. We can therefore calculate
$\langle{{ \Omega}}\rangle_g$ for any $g\leq0$ by a Monte Carlo sampling
that is free of the sign problem. If $\langle{{ \Omega} }\rangle_g$ is a
smooth function of $g$, it should then be possible to extrapolate to $g=1$
(i.e., to the original hamiltonian) from $g\leq0$. We emphasize that $g=0$ is
not expected to be a singular point of $\langle{{ \Omega} }\rangle_g$; it
is special only in the Monte Carlo evaluation.

\begin{figure} 
\label{fig_extrap} 
\epsfxsize=4truein\epsffile{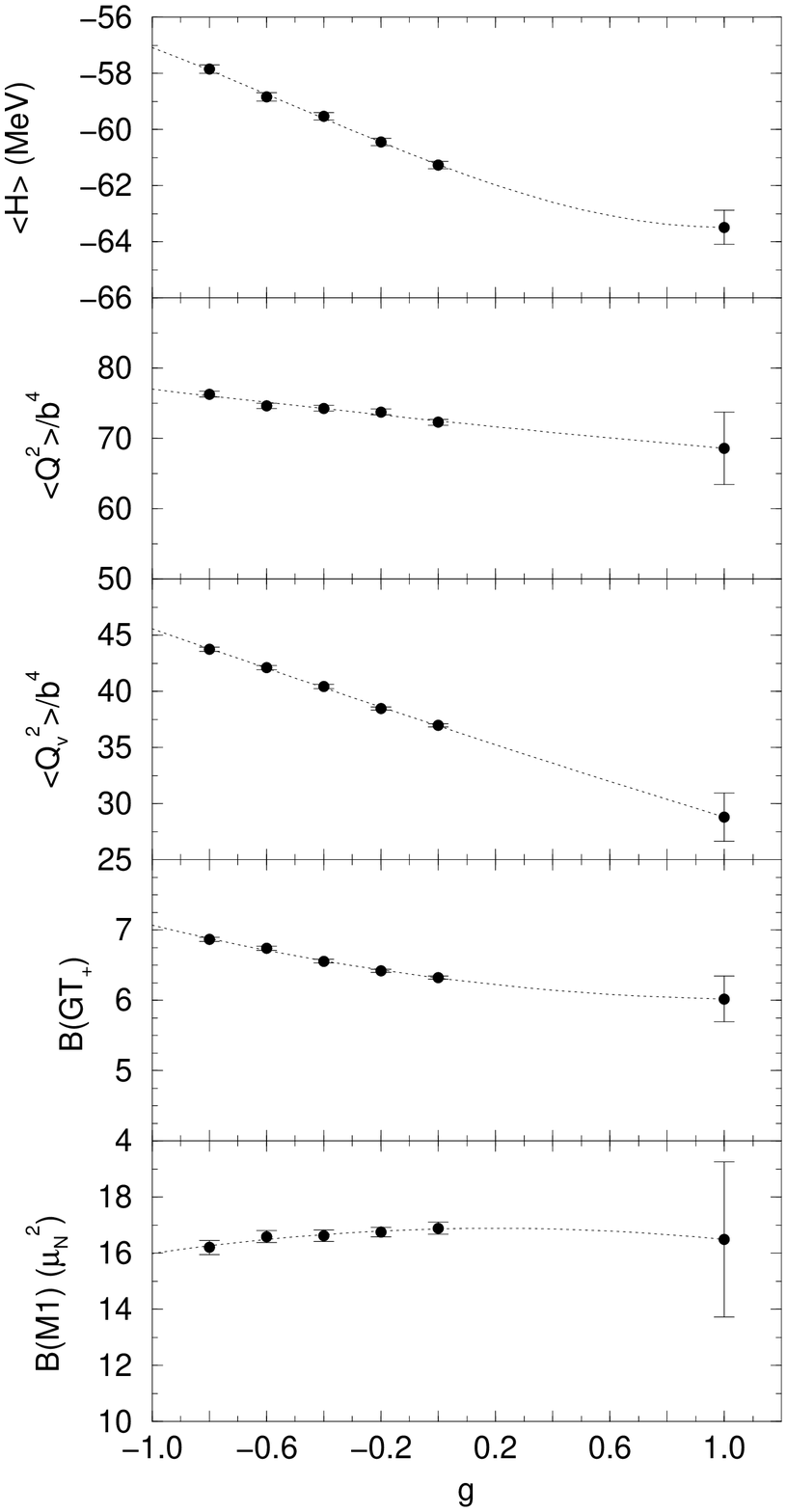} 
{Fig~\protect\ref{fig_extrap}. $g$-extrapolation of several observables for
${}^{54}$Fe calculated with the Kuo-Brown interaction KB3. } 
\end{figure}

In Fig.~\ref{fig_extrap} we exemplify the $g$-extrapolation procedure for
several observables calculated for ${}^{54}$Fe. In all cases we use
polynomial extrapolations from negative $g$-values to the physical case,
$g=1$. The degree of the polynomial is usually chosen to be the smallest that
yields a $\chi^2$ per degree of freedom less than 1. However, in several
studies, like the one of the $pf$-shell nuclei reported in Section~III, we
have conservatively chosen second-order polynomials for all extrapolations,
although in many cases a first-order polynomial already resulted in
$\chi^2$-values less than 1. At $T=0$ the variational principle requires that
$\langle  H\rangle$ has a minimum at $g=1$. We have incorporated this
fact in our extrapolations of ground state energies by using a second-order
polynomial with zero-derivative at $g=1$.

\subsection{Computational considerations}

The present SMMC code is roughly the fourth major revision of a program whose
development began in late 1990 with a single-species, single-$j$-shell
version. It is now a modular package of some 10,000 commented lines of
{\sc{FORTRAN}}. All floating point computations are double precision
(64-bit).

The package performs all of the functions necessary for shell model Monte
Carlo calculations: initialization, thermalization of the Metropolis walk,
generation of the Monte Carlo samples, evaluation of static observables and
response functions (canonical or grand-canonical), Maximum Entropy extraction
of strength functions \cite{Koonin96}, and the extrapolation in $g$ required
to solve the sign problem, as discussed in section~2.3. The data input and
results output use standard shell model conventions, so that it is easy to
change the two-body matrix elements of the interaction to incorporate
additional one- or two-body observables in the analysis, or to add or change
the orbitals in the calculation. The code has been de-bugged and tested
extensively against direct diagonalization results in the $sd$- and lower
$pf$-shells. Its operation by an experienced user can be described as
``routine,'' although it takes several weeks to acquire that experience.

Shell model Monte Carlo calculations are extraordinarily well-suited to
Multiple Instruction/Multiple Data (MIMD) architectures. Very few problems
are encountered in porting to a new machine, and the operation generally
takes less than a day. Indeed, our code is ``embarrassingly parallel'':
separate Metropolis random walks are started on each computational node,
which then produces a specified number of Monte Carlo samples at regular
intervals during the walk. Data from all of the nodes are sent to a central
node for evaluation of the Monte Carlo averages and their uncertainties.

To date, we have implemented the parallel version of our code on the Intel
DELTA and PARAGON machines at Caltech and ORNL (each with 512 i860
processors), on the 128-processor IBM SP-1 at ANL, the 512-processor IBM SP-2
at Maui, and on a Fujitsu VPP500 shared-memory vector processor (24 CPU's).
In all cases, the ratio of communications to computation is very low, with
efficiencies always greater than 95\%.

Table 1 shows benchmarks of our code on various single processors. The test
calculation involved a canonical ensemble in the full $pf$-shell ($N_s=20$
for each type of nucleon, implying $20\times20$ matrices) using a realistic
interaction. $N_t=32$ time slices were used at $\beta=2~{\rm MeV}^{-1}$
($\Delta\beta=0.0625~{\rm MeV}^{-1}$). Thirty static observables and seven
dynamical response functions were calculated at a single $g$-value (see
section 7). Note that the computational speed is independent of the
interaction and of the number of nucleons occupying the shell. Beyond using
library subroutines (BLAS and LAPACK routines), no attempt was made to
optimize the assembly level code in any of these cases. Approximately 40\% of
the computational effort is in matrix-multiply operations. 
A significant remainder of
the effort goes into building the one-body hamiltonian (13\%), setting up the
one-body evolution operators (15\%), and calculating two-body observables
(15\%), none of which is easily vectorizable. In general, the computation
time scales as $N^3_sN_t$, and is spent roughly equally on the dynamical
response functions and the static observable sampling.

\begin{table}
\begin{center}
\begin{tabular}{|c|c|c|c|}
\hline
Processor&
Peak MF&
Average MF&
Samples/hr.\\
\hline
i860 & 35 & 9 & 44 \\
IBM-SP2 Thin66& 56 & 36 & 179 \\
ALPHA-400& 56 & 28 & 141 \\
\hline
\end{tabular}
\caption{Benchmarks of the SMMC code in various processors.}
\end{center}
\end{table}

The memory required for our calculations scales as $\bar N^2_sN_t$. $\bar
N^2_s$ is the average of the squares of the numbers of neutron and proton
single-particle states. Sample values of $N_s$ for one isospin type are shown
for various model spaces in Table~2.

\begin{table} 
\begin{center} 
\begin{tabular}{|c|c|} 
\hline $0p$ &6\\ 
$1s$-$0d$ &12\\ 
$1p$-$0f$ &20\\ 
$1p$-$0f$-$0g_{9/2}$ &30\\ 
$2s$-$1d$-$0g$ &30\\ 
$2p$-$1f$-$0g_{9/2}$-$0i_{13/2}$ &44 \\ 
\hline 
\end{tabular} 
\end{center} 
\caption{Matrix dimension for various model spaces.} 
\end{table}

The code is currently structured so that a calculation with $\bar N_s=32$,
six $j$-orbitals, and $N_t=64$ time slices will fit in 12~MB of memory. A
calculation in the ($1p$-$0f$)-($2s$-$1d$-$0g$) basis has $\bar N_s=50$ and
would require about 64~MB of memory for $N_t=64$ time slices and about 128~MB
for $N_t=128$.

\section{Results}

In the past four years SMMC techniques have been applied in a variety of ways
to nuclei in the $pf$-shell and other regions using realistic or
semi-realistic interactions. In this section we will detail several of the
important results of these calculations. These are, broadly, ground state and
thermal properties of iron region nuclei, nuclear pair correlations,
$\beta\beta$-decay, and $\gamma$-soft studies.

\subsection{Ground state properties of medium mass nuclei}

\subsubsection{$pf$-shell nuclei} 
While complete $0\hbar\omega$ calculations can be carried out by direct
diagonalization in the $p$- and $sd$-shells, the exponentially increasing
number of configurations limits such studies in the next ($pf$) shell to only
the very lightest nuclei \cite{Richter,Caurier94}. SMMC techniques allow
calculation of groundstate observables in the full $0\hbar\omega$ model space
for nuclei throughout the $pf$-shell. Here, we discuss a set of such
calculations that uses the modified KB3 interaction \cite{Zuker}; the
single-particle basis is such that $N_s=20$ for both protons and neutrons.
These studies were performed for 28 even-even Ti, Cr, Fe, Ni, and Zn isotopes
and 4 odd-odd $N=Z$ nuclei. A more detailed description of the calculations
and their results are found in Ref.~\cite{Langanke95}.

\begin{figure} 
\label{fp_masses} 
$$\epsfxsize=4truein\epsffile{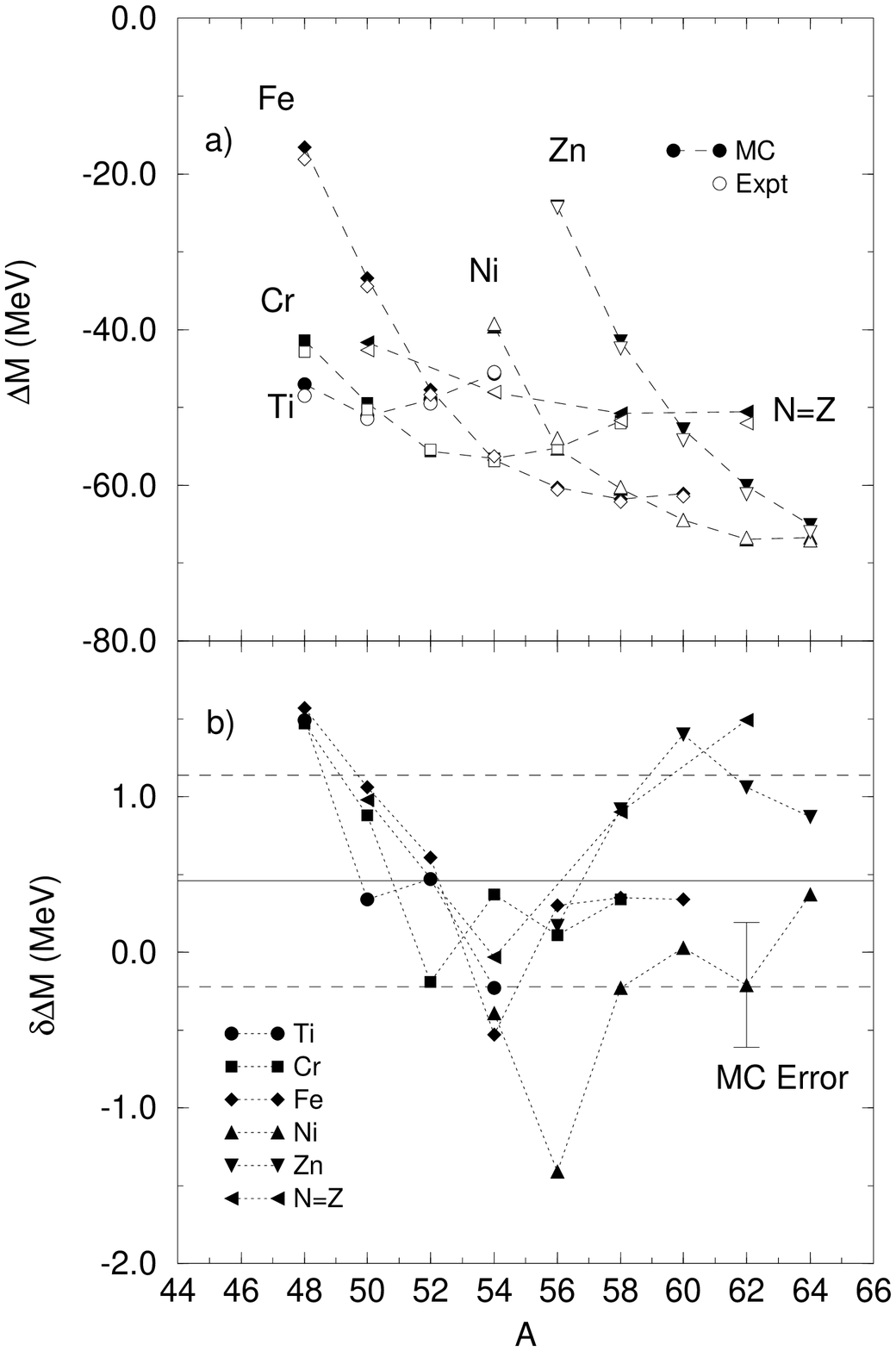}$$ 
{FIG~\protect\ref{fp_masses} Upper panel (a): Comparison of the mass excesses
$\Delta M$ as calculated within the SMMC approach with data. Lower panel (b):
Discrepancy between the SMMC results for the mass excesses and the data,
$\delta \Delta M$. The solid line shows the average discrepancy, 450 keV,
while the dashed lines show the rms variation about this value (from
\protect\cite{Langanke95}). } 
\end{figure}

Figure \ref{fp_masses} shows systematic results for the mass defects
obtained directly from $\langle  H\rangle$. The SMMC results have been
corrected for the coulomb energy, which is not included in the KB3
interaction, using \cite{Caurier94} 
\begin{equation} 
H_{\rm Coul} = {\pi(\pi-1)\over2} \cdot 0.35 - \pi \nu 0.05 + \pi \cdot\;,
7.289. \label{hcoul} 
\end{equation} 
where $\pi$ and $\nu$ are the numbers of valence protons and neutrons,
respectively, and the energy is in MeV. As in Ref.~\cite{Caurier94}, we have
increased the calculated energy expectation values by $0.014 \cdot
n(n-1)$~MeV (where $n=\pi+\nu$ is the number of valence nucleons) to correct
for a ``tiny'' residual monopole defect in the KB3 interaction. In general,
there is excellent agreement; the average error for the nuclei shown is
$+0.45$~MeV, which agrees roughly with the internal excitation energy of a
few hundred keV expected in our finite-temperature calculation.

\begin{figure} 
\label{fp_be2} 
$$\epsfxsize=4truein\epsffile{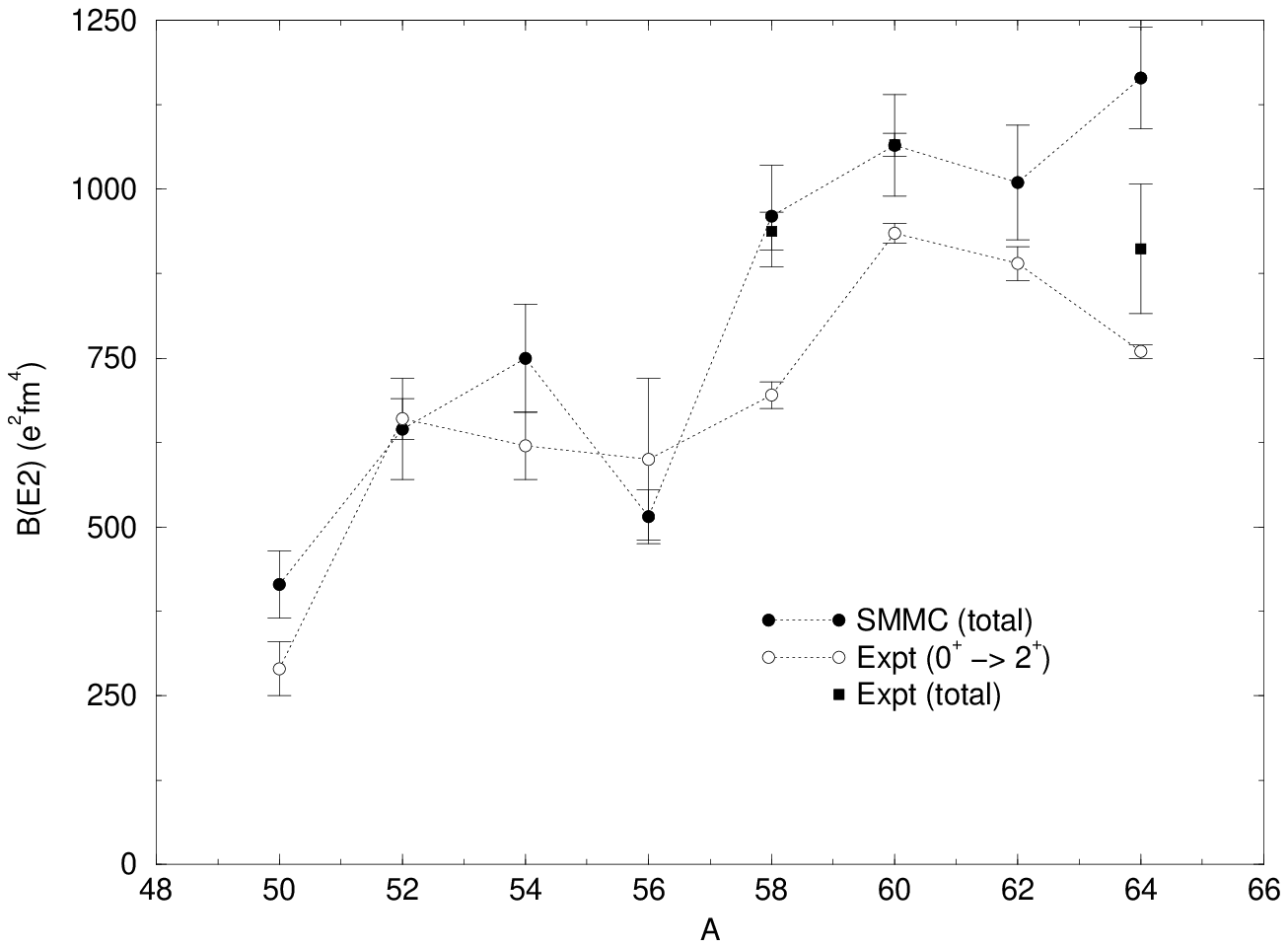}$$ 
{FIG~\protect\ref{fp_be2} Comparison of the experimental $B(E2, 0_1^+
\rightarrow 2_1^+)$ strengths with the total $B(E2)$ strength calculated in
the SMMC approach for various $pf$-shell nuclei having either a proton or
neutron number of $28$. For the nickel isotopes ${}^{58,60,64}$Ni the total
$B(E2)$ strength (full squares) is known from inelastic electron scattering
data (from \protect\cite{Langanke95}). } 
\end{figure}

Figure \ref{fp_be2} shows the calculated total $E2$ strengths for selected
$pf$-shell nuclei. This quantity is defined as 
\begin{equation} 
B(E2) = \langle (e_p { Q}_p + e_n { Q}_n )^2 \rangle \; , 
\end{equation} 
with 
\begin{equation} 
{ Q}_{p(n)} = \sum_i r_i^2 Y_2 (\theta_i, \phi_i) \; , 
\end{equation} 
where the sum runs over all valence protons (neutrons). The effective charges
were chosen to be $e_p=1.35$ and $e_n=0.35$, while we used $b=1.01 A^{1/6}$
fm for the oscillator length. Shown for comparison are the $B(E2)$ values for
the $0^+_1\rightarrow 2^+_1$ transition in each nucleus; in even-even nuclei
some 20-30\% of the strength comes from higher transitions. The overall trend
is well reproduced. For the nickel isotopes ${}^{58,60,64}$Ni, the total
$B(E2)$ strength is known from $(e,e')$ data and agrees very nicely with our
SMMC results.

\subsubsection{The $0f_{5/2}1p0g_{9/2}$ model space}

In order to investigate heavier systems, the $0g_{9/2}$ was included in the
$pf$ model space\cite{Kuo}. However, we found that the coupling between the
$0f_{7/2}$ and $0g_{9/2}$ orbitals causes significant center-of-mass
contamination to the ground state, and we therefore close the $0f_{7/2}$. The
model space is thus $0f_{5/2}1p0g_{9/2}$. This appears to be a good
approximation in systems where $N$ and $Z$ are greater than 28. 
The monopole terms of this new interaction were modified
\cite{Nowacki} to give a good description of the spectra of nuclei in the Ni
isotopes. Since ${}^{56}$Ni is the core of this model space, the
single-particle energies were determined from the ${}^{57}$Ni spectrum. 

Shell-model masses must be corrected for coulomb effects and for
`grand-monopole' terms \cite{Dean97} that are not taken into account when 
determining interaction matrix elements from nuclear spectra. 
Following Ref.~\cite{Caurier94} we have fitted a correction to the 
calculated binding energies which takes into
account the coulomb energy and residual effects of the monopole terms. 
The form of this correction is 
\begin{equation}
H_c=\alpha \pi (\pi-1)+\beta\pi\nu+\gamma\pi+\delta n(n-1) +\varepsilon n
\end{equation}
where $\pi,\nu$ are the number of valence protons and
neutrons, and $n=\pi+\nu$, and all parameters are in units
of MeV. $\gamma$ is held fixed at 7.289. 
Minimizing $\chi^2$ for the difference between experimental and theoretical
binding energies relative to the $^{56}$Ni core gives
$\alpha=0.234$, $\beta=0.0156$, $\delta=0.0316$, and $\varepsilon=1.828$.

\begin{figure} 
\label{Fig2} 
\begin{center} 
$$\epsfxsize=4truein\epsffile{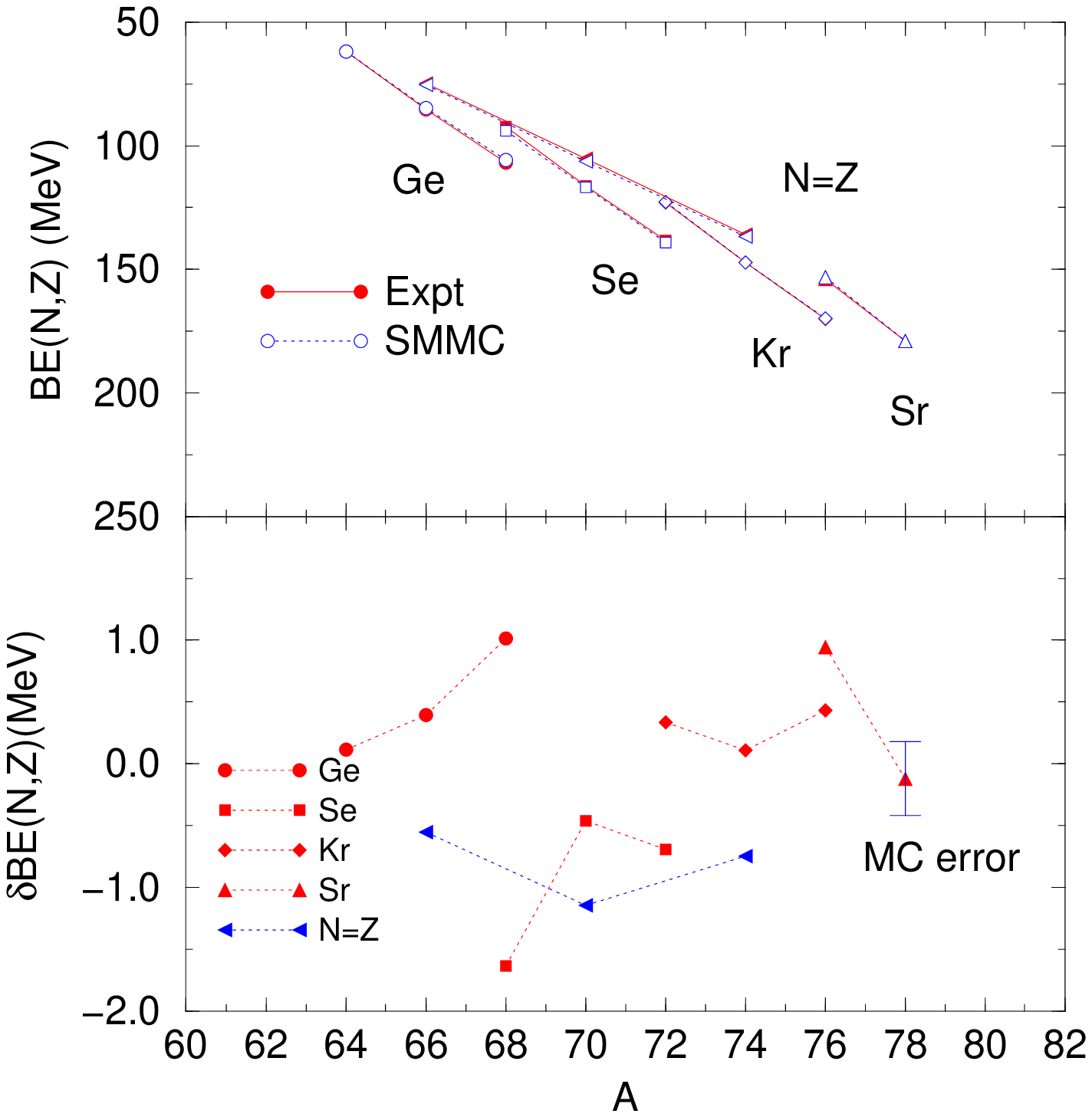}$$ 
\end{center} 
{FIG~\protect\ref{Fig2}. Calculated mass excesses for nuclei in the mass
range $A=64$--80 are compared with experiment (top panel), and the difference
between experiment and theory is shown (bottom panel). } 
\end{figure}

Shown in Fig.~\ref{Fig2} are the experimental and theoretical binding energies
$BE(N,Z)$ for the nuclei studied here (top panel), and the difference
between experiment and theory (bottom panel). The overall agreement
is reasonable.

\subsubsection{Gamow-Teller strengths and distributions}

The Gamow-Teller (GT) properties of nuclei in this region of the periodic
table are crucial for supernova physics \cite{Bethe90}. The core of a massive
star at the end of hydrostatic burning is stabilized by electron degeneracy
pressure as long as its mass does not exceed the appropriate Chandrasekhar
mass $M_{CH}$. If the core mass exceeds $M_{CH}$, electrons are captured by
nuclei \cite{Bethe90}. For many of the nuclei that determine the electron
capture rate in this early stage of the presupernova \cite{Aufderheide},
Gamow-Teller (GT) transitions contribute significantly to the electron
capture rate. Due to insufficient experimental information, the GT$_+$
transition rates have so far been treated only qualitatively in presupernova
collapse simulations, assuming the GT$_+$ strength to reside in a single
resonance whose energy relative to the daughter ground state has been
parametrized phenomenologically \cite{FFN}; the total GT$_+$ strength has
been taken from the single particle model. Recent $(n,p)$ experiments
\cite{gtdata1}-\cite{gtdata5}, however, show that the GT$_+$ strength is
fragmented over many states, while the total strength is significantly
quenched compared to the single particle model. (A recent update of the
GT$_+$ rates for use in supernova simulations assumed a constant quenching
factor of 2 \cite{Aufderheide}.)

\begin{figure} 
\label{fig_2} 
\epsfxsize=3.75truein\epsffile{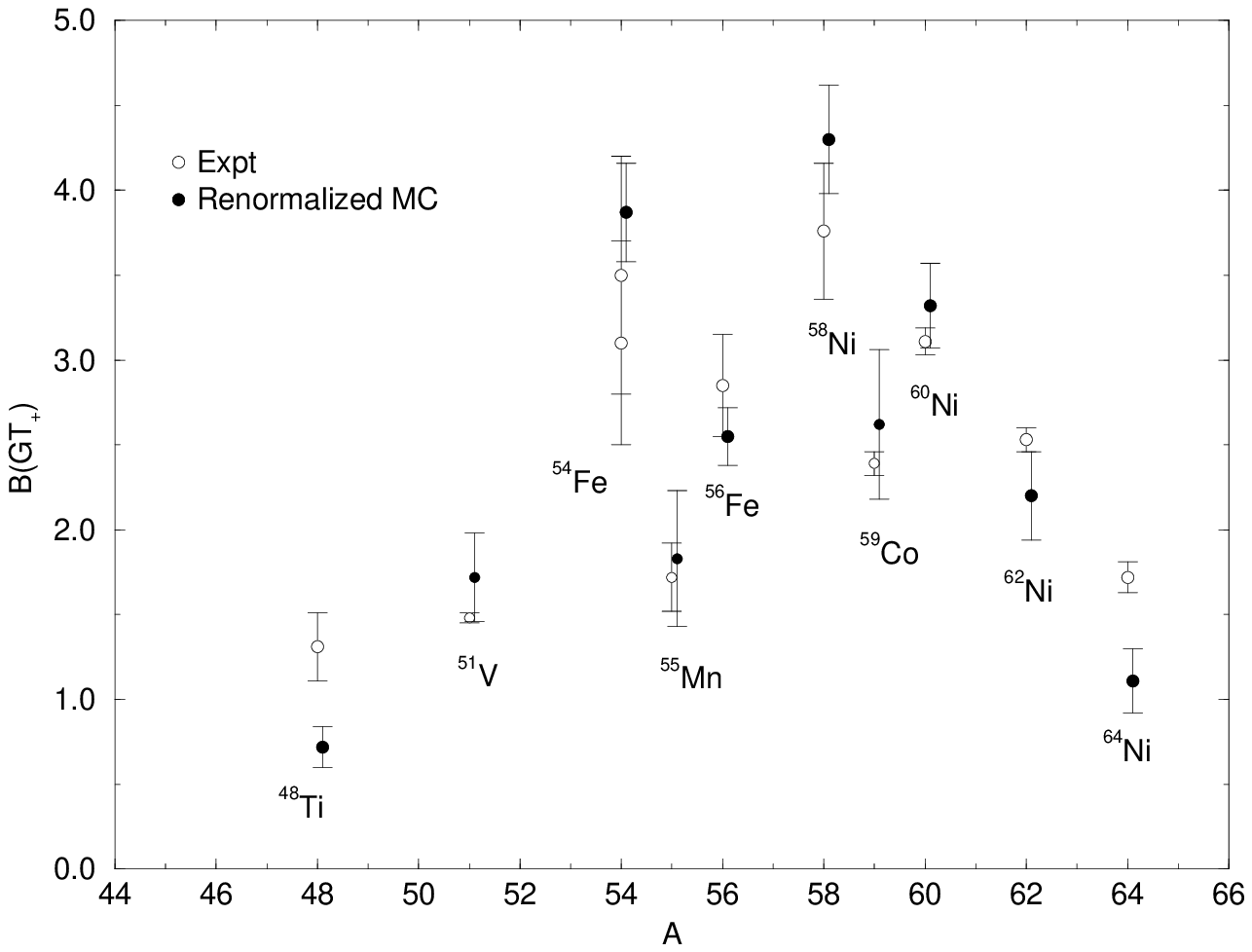} 
{FIG~\protect\ref{fig_2}. Comparison of the renormalized total Gamow-Teller
strength, as calculated within the present SMMC approach, and the
experimental $B(GT_+)$ values deduced from $(n,p)$ data
\protect\cite{gtdata1}-\protect\cite{gtdata5}. [For the odd nuclei
(${}^{51}$V, ${}^{55}$Mn and ${}^{59}$Co) the SMMC calculations have been
performed at $\beta =1$ MeV$^{-1}$ to avoid the odd-$A$ sign problem. For
even-even nuclei the GT strength calculated at this temperature is still a
good approximation for the ground state value, as discussed in the text.]}
\end{figure}

The total GT strengths are defined as 
\begin{equation} 
B(GT_{\pm}) = \langle ({\vec \sigma} \tau_{\pm} )^2 \rangle . 
\end{equation} 
From $0 \hbar \omega$ shell model studies of the GT strengths for $sd$ shell
nuclei and for light $pf$ shell nuclei it has been deduced that the spin
operator in (31) should be replaced by an effective one, ${\vec \sigma_{\rm
eff}} = {\vec \sigma } / 1.26$ \cite{Wild84,Caurier94}. This renormalization
is not well understood. It is believed to be related either to a second-order
core polarization caused by the tensor force \cite{Hamomoto}, or to the
screening of the Gamow-Teller operator by $\Delta$-hole pairs \cite{Delta}.
Using the effective spin operator, our calculated $B({\rm GT}_+)$ are in
excellent agreement with the data deduced from $(n,p)$ reactions
(Figure~\ref{fig_2}).  Thus, our calculations support both the statement that
the spin operator is renormalized by a universal factor $(1/1.26)$ in nuclei,
and the statement that complete shell-model calculations can account for the
GT strength observed experimentally.

In a series of truncated shell-model calculations, Aufderheide and
collaborators have demonstrated that a strong phase space dependence makes
the Gamow-Teller contributions to the presupernova electron capture rates
more sensitive to the strength {\it distribution} in the daughter nucleus
than to the total strength \cite{Aufder1}. In this work it also became
apparent that complete $0 \hbar \omega$ studies of the GT$_+$ strength
distribution are desirable. Such studies are now possible using the SMMC
approach.

To determine the GT$_+$ strength distribution, we have calculated the response
function of the ${\bf \sigma} \tau_+$ operator, $R_{\rm GT} (\tau)$, as
defined in Eq.~(\ref{eq_e}). As the strength function $S_{\rm GT}(E)$ is the
inverse Laplace transform of $R_{\rm GT} (\tau)$, we have used the Maximum
Entropy technique, described in Ref.~\cite{Koonin96}, to extract $S_{\rm
GT}(E)$.

\begin{figure} 
\label{gt_resp} 
$$\epsfxsize=5truein\epsffile{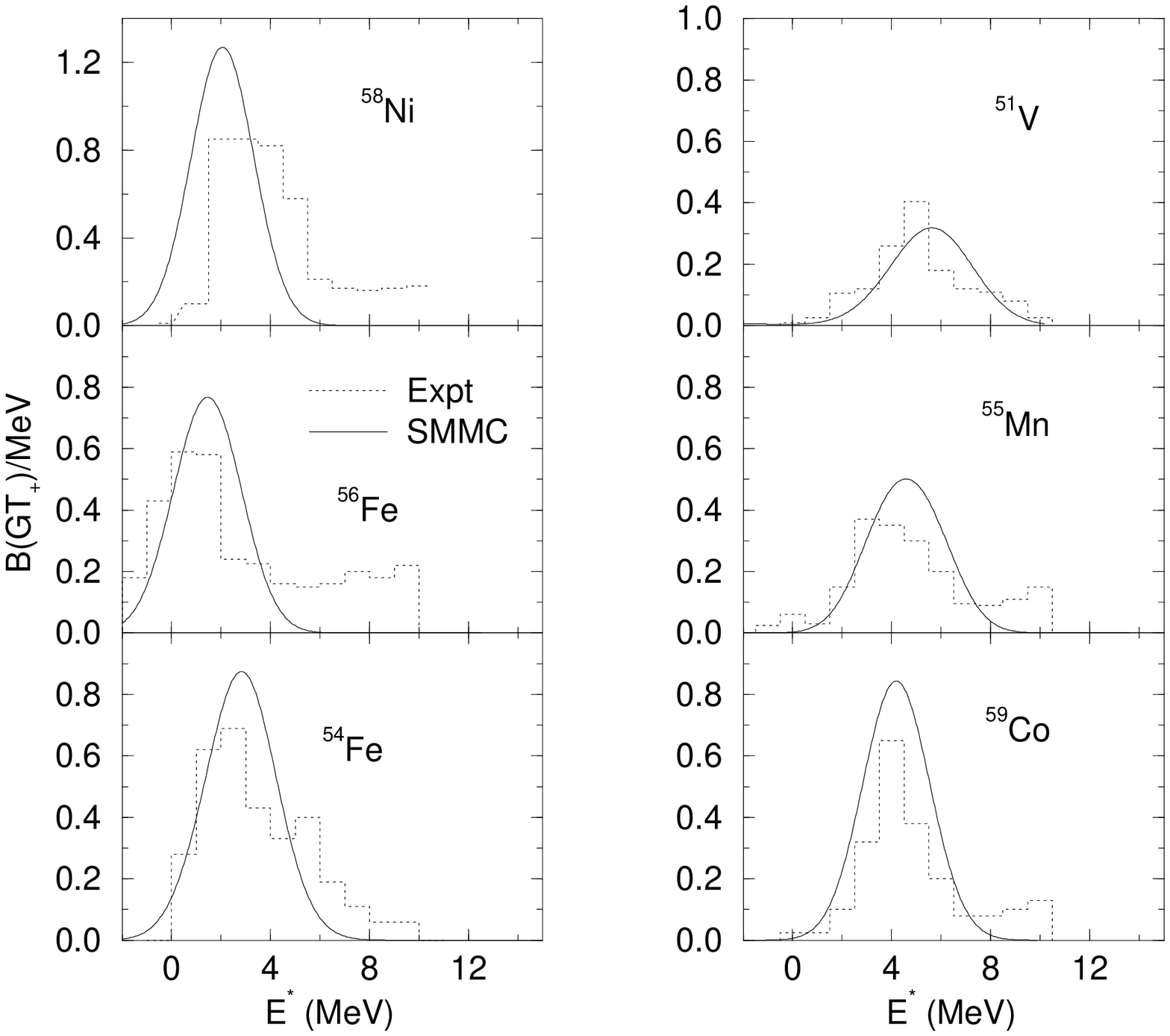}$$ 
\protect\label{gtdata2}. {FIG~\protect\ref{gt_resp}. Shown are
SMMC GT$_+$ strength
distributions (solid line) for various nuclei in the iron region. 
The energies refer to the daughter nuclei. The dashed histograms show
the experimental strength distribution as extracted from $(n,p)$ data.
(Following \protect\cite{Poves} the calculated strength distributions have
been folded with Gaussians of width 1.77~MeV to account for the experimental
resolution.) } 
\end{figure}

As first examples we have studied several nuclei (${}^{51}$V, ${}^{54,56}$Fe,
${}^{55}$Mn, ${}^{58,60,62,64}$Ni, and ${}^{59}$Co), for which the
Gamow-Teller strength distribution in the daughter nucleus is known from
$(n,p)$ experiments \cite{gtdata1}-\cite{gtdata5}. However, note that
electron capture by these nuclei plays only a minor role in the presupernova
collapse. As SMMC calculates the strength function within the parent nucleus,
the results have been shifted using the experimental $Q$-values, and 
the Coulomb
correction has been performed using Eq.~(\ref{hcoul}). For all nuclei, the
SMMC approach calculates the centroid and width of the strength distribution
in good agreement with data (see Fig.~\ref{gt_resp}). The centroid of the
GT$_+$ strength distributions is found to be nearly independent of
temperature, while its width increases with temperature.

Following the formalism described in Refs.~\cite{FFN,Aufderheide}, the
Gamow-Teller contributions to the electron capture rates under typical
presupernova conditions have been calculated, assuming that the electrons have
a Fermi-Dirac distribution with a chemical potential adopted from the stellar
trajectory at the electron-to-nucleon ratio corresponding to the particular
nucleus \cite{Aufderheide}. The calculations have been performed using both
the SMMC and experimental GT$_+$ strength distributions
\cite{gtdata1}-\cite{gtdata5}. The two electron capture rates agree within a
factor of two for temperatures $T=(3-5) \times 10^9$ Kelvin, which is the
relevant temperature regime in the presupernova collapse \cite{Aufderheide}.
Thus, for the first time it is possible to calculate, with reasonable
accuracy, the electron capture rate for nuclei like ${}^{55}$Co or
${}^{56}$Ni, which dominate the electron capture process in the early
presupernova collapse \cite{Aufderheide}.

\subsection{Pair correlations}

The residual nuclear interaction builds up pairing correlations in a nucleus.
Introducing nucleon creation operators $a^\dagger$, these correlations can be
studied by defining pair creation operators 
\begin{equation} 
A^\dagger_{JM}(j_aj_b) = \frac{1}{\sqrt{1+\delta_{ab}}} 
\left[a^\dagger_{j_a}\times a^{\dagger}_{j_b} \right]_{JM} 
\end{equation} 
for proton-proton or neutron-neutron pairs, and 
$$ A^\dagger_{JM}(j_aj_b) = 
\frac{1}{\sqrt{2(1+\delta_{ab})}} \left\{\left[a^\dagger_{pj_a}\times 
a^{\dagger}_{nj_b} \right]_{JM} \right. $$ 
\begin{equation} \;\;\;\;\;\;\;\;\;\;\;\; \left. \pm
\left[a^\dagger_{nj_a}\times a^{\dagger}_{pj_b} \right]_{JM} \right\} \;,
\label{maa} 
\end{equation} 
for proton-neutron pairs where ``$+(-)$" is for $T=0 (T=1)$ $pn$-pairing.
With these definitions, we construct a pair matrix 
\begin{equation} 
M_{\alpha \alpha'}^J = 
\sum_M \langle A_{JM}^\dagger (j_a,j_b) A_{JM} (j_c,j_d) \rangle\; ,
\label{pair_a} 
\end{equation} 
where $\alpha=\left\{j_a,j_b\right\}$ and $\alpha'=\left\{j_c,j_d\right\}$
and the expectation value is in the ground state or canonical ensemble at a
prescribed temperature. The pairing strength for a given $J$ is then given by
\begin{equation} 
P(J)=\sum_{\alpha\ge\alpha'}M_{\alpha,\alpha'}^J. \label{pair_b} 
\end{equation}

An alternative measure of the overall pair correlations is given in terms of
the BCS pair operator 
\begin{equation} 
\Delta_{JM}^\dagger = \sum_{\alpha} A_{JM}^\dagger (\alpha). 
\end{equation} 
The quantity $\sum_M \langle \Delta_{JM}^\dagger \Delta_{JM} \rangle$ is then
a measure of the number of nucleon pairs with spin $J$. For the results
discussed below, the BCS-like definition for the overall pairing strength
yields the same qualitative results for the pairing content as the definition
(\ref{pair_b}). Some SMMC results for BCS pairing in nuclei $A=48-60$ are
published in Refs.~\cite{Langanke95,Langanke96,Langanke,Engel}.

With our definition in (\ref{pair_b}) the pairing strength is positive at the
mean-field level. The mean-field pairing strength, $P_{\rm MF} (J)$, can be
defined by replacing the expectation values of the two-body matrix elements
in the definition of $M^J$ by 
\begin{equation} \langle a_1^\dagger
a_2^\dagger a_3 a_4 \rangle \rightarrow n_1 n_2 \left( \delta_{13}
\delta_{24} - \delta_{23} \delta_{14} \right) , \label{pair_c} 
\end{equation}
where $n_k = \langle a_k^\dagger a_k \rangle$ is the occupation number of the
orbital $k$. This mean-field value provides a baseline against which true
pair correlations can be judged, by defining
\begin{equation} 
P_{\rm cor}(J)=P(J) - P_{\rm MF} (J) \;. \label{pair_d} 
\end{equation}

\subsubsection{Ground state pair correlations}

In a first project we studied the pair correlations in several even-even $pf$
shell nuclei ($^{54,56,58}$Fe and $^{56}$Cr) \cite{Langanke96}. As expected,
we found a large excess of $J=0^+$ like-particle pairing in the ground states
of these nuclei. With increasing temperature, these pairing correlations
decrease and at around $T=1$ MeV the like-particle pairs break in these
nuclei (pairing phase transition). Our calculations also indicate that
isoscalar proton-neutron (mainly $J=1^+$) pairs persist to higher
temperatures. In Ref.~\cite{Langanke96} we have related the thermal
dependence of several observables to the temperature dependence of associated
pairing correlations.

It has long been anticipated that $J=0^+$ proton-neutron correlations play an
important role in the ground states of $N=Z$ nuclei. These correlations were
explored with SMMC for $N=Z$ nuclei with $A=48-58$ in the $pf$-shell
\cite{Langanke}, and $A=64-74$ in the $0f_{5/2}1p0g_{9/2}$ space. As the
even-even $N=Z$ nuclei have isospin $T=0$, $\langle A^\dagger A\rangle$ is
identical in all three isovector $0^+$ pairing channels. This symmetry does
not hold for the odd-odd $N=Z$ nuclei in this mass region, which usually have
$T=1$ ground states, and $\langle {A}^\dagger {A} \rangle$ can differ
for proton-neutron and like-nucleon pairs. (The expectation values for proton
pairs and neutron pairs are identical.)

\begin{figure} 
\label{Fig3} 
\begin{center} 
$$\epsfxsize=3.4truein\epsffile{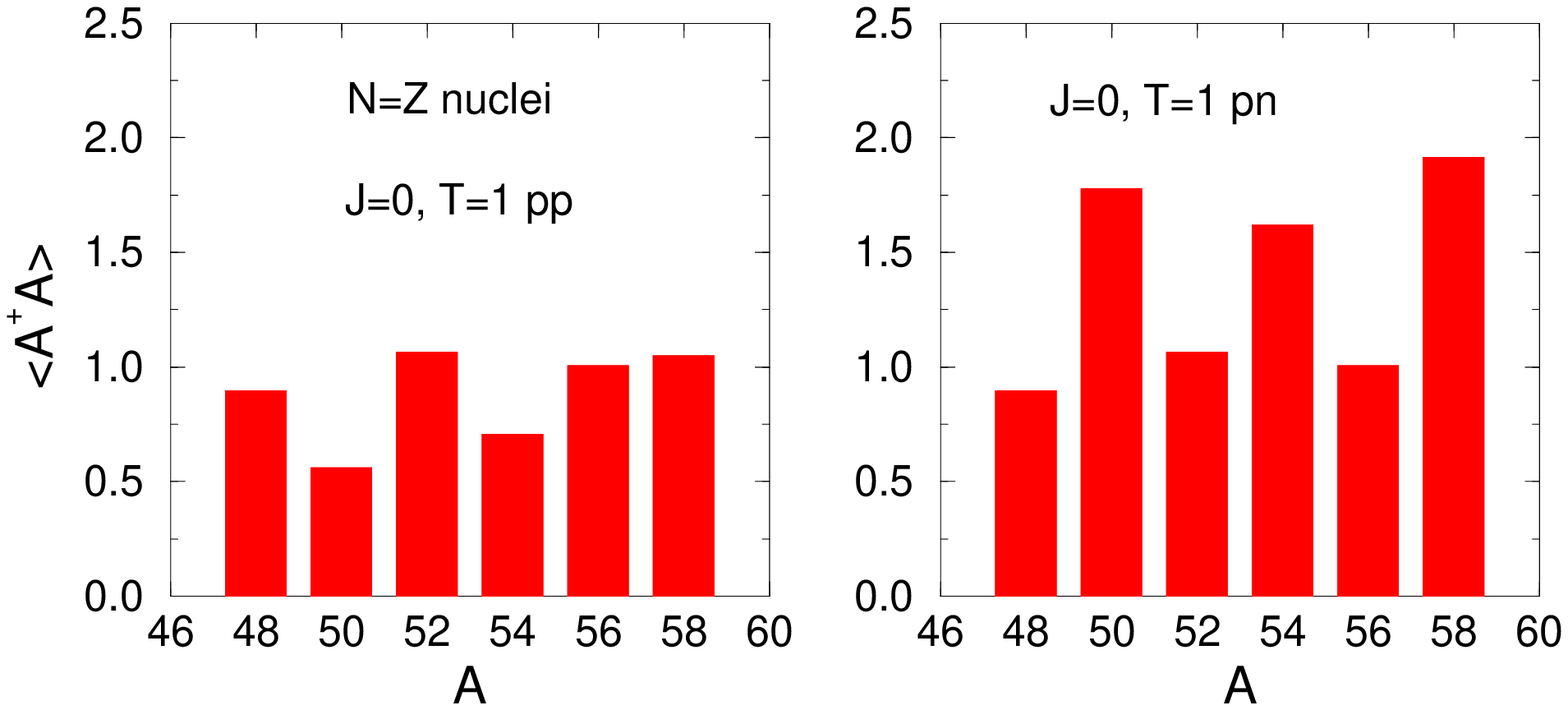}$$ 
\end{center} 
{Fig~\protect\ref{Fig3}. Largest eigenvalues for the $J=0$, $T=1$
proton-proton (left) and proton-neutron (right) pairing matrix as a function
of mass number. } 
\end{figure}

We find the proton-neutron pairing strength is significantly larger for odd-odd
$N=Z$ nuclei than in even-even nuclei, while the $0^+$ proton and neutron
pairing shows the opposite behavior, in both cases leading to a noticeable
odd-even staggering, as displayed in Fig.~\ref{Fig3} for the $pf$ shell. Due
to the strong pairing in the $f_{7/2}$ orbital, all three isovector $0^+$
channels of the pairing matrix exhibit essentially only one large eigenvalue
which is used as a convenient measure of the pairing strength in
Fig.~\ref{Fig3}. This staggering is caused by a constructive (destructive)
interference of the isotensor and isoscalar parts of ${A}^\dagger
{A}$ in the odd-odd (even-even) $N=Z$ nuclei. The isoscalar part is
related to the pairing energy and is roughly constant for the nuclei we have
studied.

\begin{figure} 
\label{Fig4} 
\begin{center} 
$$\epsfxsize=3.5truein\epsffile{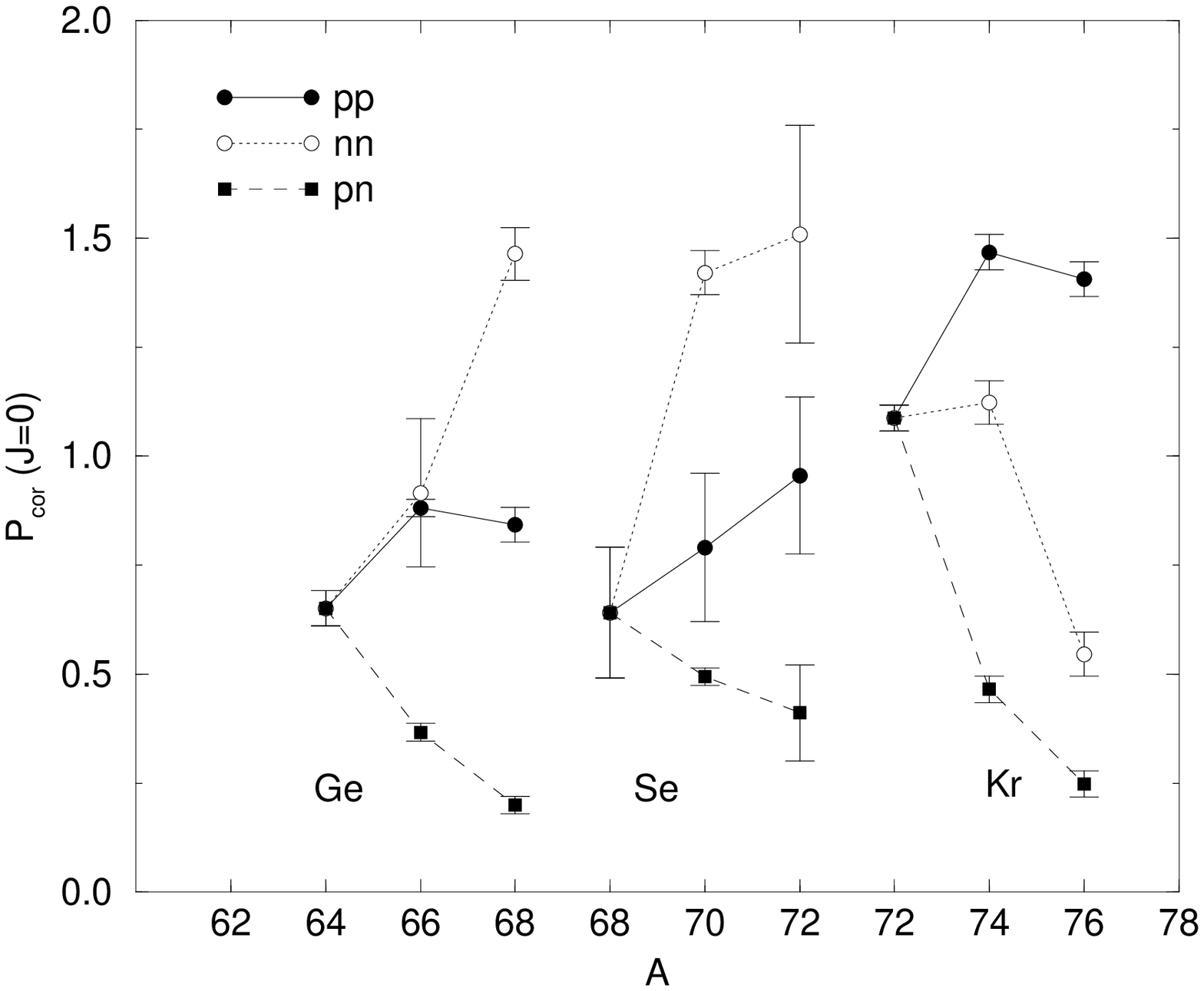}$$ 
\end{center} 
{Fig~\protect\ref{Fig4}. Correlated $J=0$, $T=1$ pairs in the proton-proton
and proton-neutron channels for selected isotope chains in the
$0f_{5/2}1p0g_{9/2}$ model space. } 
\end{figure}

Fig.~\ref{Fig4} shows the correlated pairs for $N=Z$ nuclei in the $A=64-74$
region of the $0f_{5/2}1p0g_{9/2}$ space. 
The correlated pairs exhibit a strong $J=0$, $T=1$ like-particle staggering
for the even-even and odd-odd $N=Z$ systems, while the number of correlated
proton-neutron pairs is much larger than the like-particle number for the
odd-odd systems. The correlated pairing behavior of $N=Z$, $N=Z+2$, $N=Z+4$
nuclei is shown in Fig.~\ref{Fig4}, where one clearly sees the decrease in
T=1 proton-neutron pairing as one moves away from $N=Z$. As discussed by
Engel {\it et al.} \cite{Engel}, increasing the number of neutron pairs
increases the collectivity of the neutron condensate, making fewer neutrons
available to pair with protons. As a result, the protons pair more often with
one another, although their number has not changed, and the np pairing drops
drastically. A weak neutron shell closure is evident in the Kr isotopes as
the neutrons fill to N=40 at $^{76}$Kr. Although the neutron
pairing correlations decrease here, they are not zero as the $g_{9/2}$ 
has some occupation. 

\subsubsection{Pairing and rotation}

\begin{figure} 
\label{rbrot_fig} 
\begin{center} 
$$\epsfxsize=4.5truein\epsffile{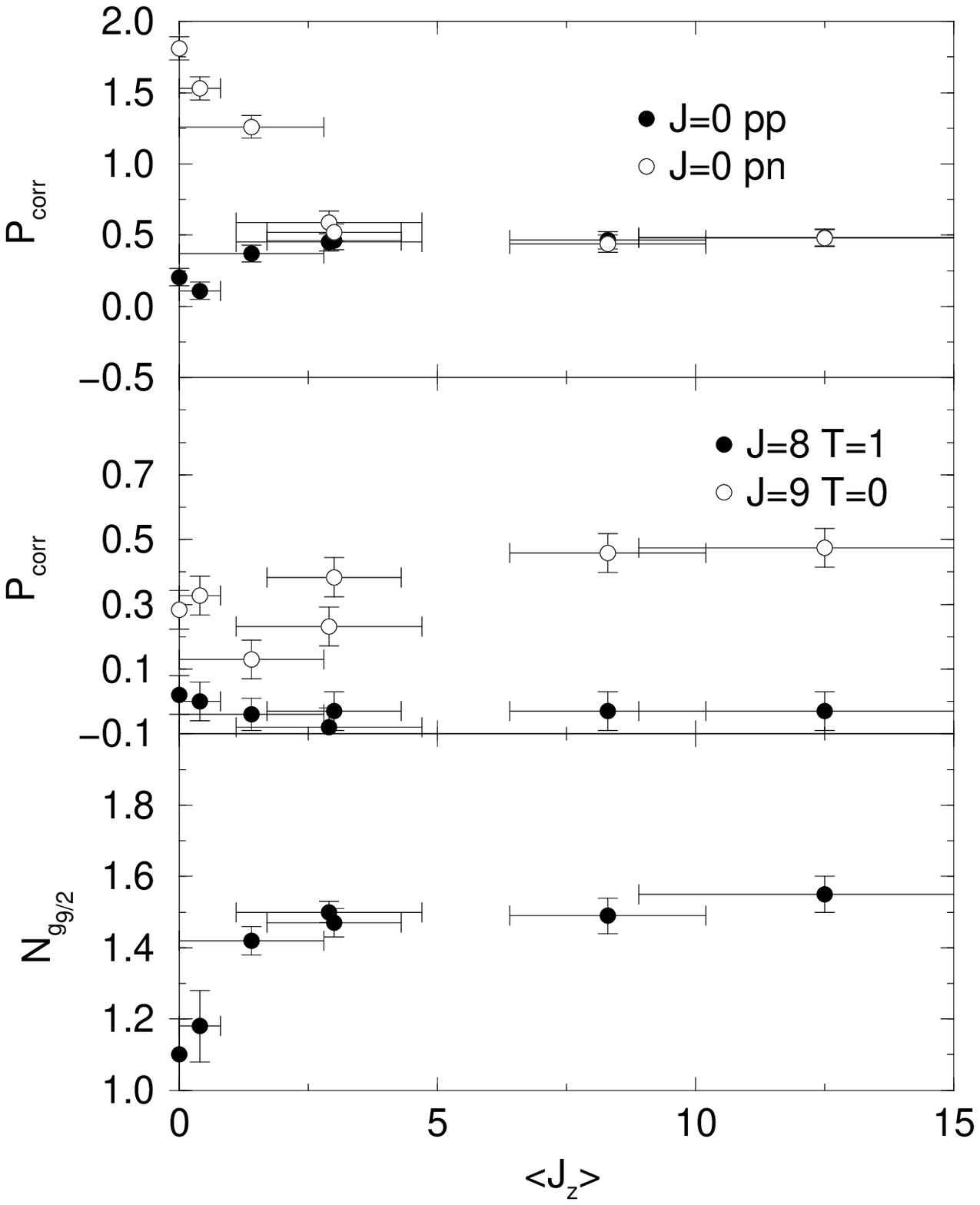}$$ 
\end{center} 
{Fig~\protect\ref{rbrot_fig}. Selected pair correlations and the proton
$g_{9/2}$ occupation number (bottom panel) as a function of $\langle J_z
\rangle$. The top panel shows the isovector $J=0$ $pp$ and $pn$ correlations,
while the isoscalar $J=9$ and isovector $J=8$ $pn$ correlations are shown in
the middle panel. } 
\end{figure}

A recent study \cite{rb_expt}of the $\gamma$ decays of the odd-odd $N=Z$
nucleus $^{74}$Rb revealed the phenomenon of an isospin band crossing at
modest excitation energies. While the ground state rotational band can be
identified as being formed from the $T=1$ isobaric analogue states of
$^{74}$Kr, a $T=0$ band becomes energetically favored over the $T=1$ band
with increasing rotational frequency. To study this isospin band crossing, we
have performed a cranked SMMC calculation \cite{dean96} in which the shell
model hamiltonian is replaced by $H\rightarrow H+ \omega J_z$. Note that
since $J_z$ is a time-odd operator, the sign problem is reintroduced, and
good statistical sampling imposes a limit on $\omega$. For the calculations
presented in Table~3, the largest cranking frequency was $\omega=0.4$~MeV. In
agreement with experiment, we find $T=1$ for the ground state ($\omega = 0$).
However, with increasing frequency $T=0$ states are mixed in (and even
dominate at large angular momenta) and $\langle T^2\rangle$ decreases, as can
be seen in Table~3. Thus our calculation confirms that $^{74}$Rb changes from
a $T=1$ dominated system to one dominated by $T=0$ states with increasing
rotational frequency. To understand the apparent crossing of the $T=1$ and
$T=0$ bands, we have studied the various pair correlations as a function of
rotational frequency. We find that the isovector $J=0$ correlations and the
aligned isoscalar $J=9$ $pn$ correlations are most important in this
transition. These correlations are plotted in Fig.~\ref{rbrot_fig} as a
function of $\langle J_z \rangle$, where we have defined pair correlations by
Eq.~(\ref{pair_d}). Strikingly, the largest pair correlations are found in the
isovector $J=0$ and isoscalar $J=9$ proton-neutron channels at low and high
frequencies, respectively. Furthermore, the variation of isospin with
increasing frequency reflects the relative strengths of these two $pn$
correlations. Our calculation clearly confirms that proton-neutron
correlations determine the behavior of the odd-odd $N=Z$ nucleus $^{74}$Rb,
as already supposed in \cite{rb_expt}.

With increasing frequency, the isovector $J=0$ $pn$ correlations decrease
rapidly to a constant at $\langle J_z \rangle \approx 3$. This behavior is
accompanied by an increase of the $pn$ correlations in the maximally aligned
channel, $J=9,T=0$ which dominates $^{74}$Rb at rotational frequencies where
the $J=0$ $pn$ correlations become small. Furthermore, although $J=8$, $T=1$
pairs exist, they do not exhibit correlations beyond the mean field, as shown
in Fig.~\ref{rbrot_fig}. We note that the experiment also indicates that the
mixing of $T=0$ states sets in near $J=3$.

\begin{table} 
\begin{center} 
\caption{Summary of various properties of ${}^{74}$Rb. These results, quoted
as a function of the cranking frequency $\omega$, include
$\protect\sqrt{J^2}$, $\langle T^2 \rangle $, the proton-proton $J=0$ pairing
correlations ($J=0$, $pp$), $J=0$, $T=1$, $pn$ pairing correlations ($J=0$,
$pn$). } 
\begin{tabular}{|c|c|c|c|c|} 
$\omega$~MeV & $\sqrt{\langle J^2 \rangle}$ & $\langle T^2 \rangle $ &$J=0$
$pp$ & $J=0$ $pn$ \\ 
\hline 0.00 & 3.5 $\pm$ 0.6 & 1.97 $\pm$ 0.05 & 0.21 $\pm$ 0.05 &1.81$\pm$
0.08 \\ 
0.10 & 2.1 $\pm$ 0.9 & 1.825$\pm$ 0.03 & 0.11 $\pm$ 0.05 &1.53$\pm$0.08 \\ 
0.20 & 2.9 $\pm$ 0.7 & 1.98 $\pm$ 0.10 & 0.37 $\pm$ 0.06 &1.26$\pm$0.08 \\ 
0.25 & 2.6 $\pm$ 1.1 & 1.72 $\pm$ 0.16 & 0.45 $\pm$ 0.06 &0.60$\pm$0.08 \\ 
0.30 & 3.3 $\pm$ 0.8 & 1.77 $\pm$ 0.13 & 0.46 $\pm$ 0.06 &0.52$\pm$0.08 \\ 
0.35 & 6.6 $\pm$ 3.1 & 1.40 $\pm$ 0.14 & 0.46 $\pm$ 0.06 &0.44$\pm$0.08 \\ 
0.40 & 11.3 $\pm$ 1.9 & 0.69 $\pm$ 0.20 & 0.48 $\pm$ 0.06 &0.48$\pm$0.08 \\ 
\hline
\end{tabular} 
\end{center} 
\end{table}

\subsubsection{Pairing and temperature}

Another striking difference in proton-neutron pairing can be found in the
thermal properties of odd-odd (e.g. ${}^{50}$Mn) and even-even (e.g.
${}^{52}$Fe) $N=Z$ nuclei \cite{Langanke}. As in other even-even $N=Z$
nuclei, the ground state isospin of $^{52}$Fe is $T=0$ and isospin symmetry
forces $pp$, $nn$, and $np$ pairing to be identical. With increasing
temperature, $T=1$ components are slowly mixed in, breaking the symmetry
between $nn$-$pp$ and $np$ pairing. However, $\langle T^2 \rangle \approx 0$
for $T<1$~MeV and the symmetry holds. As $J=0^+$ pairs break at around this
temperature in even-even nuclei in this mass range, the phase transition is
clearly noticeable in all three isovector pairing channels.

As discussed above, the ground state of the odd-odd nucleus ${}^{50}$Mn (with
isospin $T=1$) is dominated by $0^+$ $pn$ correlations. As a striking feature
shown in Fig.~\ref{mn50}, the proton-neutron pairing decreases rapidly with
temperature and has dropped to the mean-field value by $T\sim 0.75$~MeV,
while the like-particle pairing remains roughly constant to 1.1~MeV. The
vanishing of the $pn$ correlations is accompanied by a change in isospin,
which decreases from the ground state value $\langle T^2 \rangle =2$ to
$\langle T^2 \rangle \approx0.2$ at temperatures near $T=1$~MeV.

\begin{figure} 
\label{fe52} 
\begin{center} 
$$\epsfxsize=4.5truein\epsffile{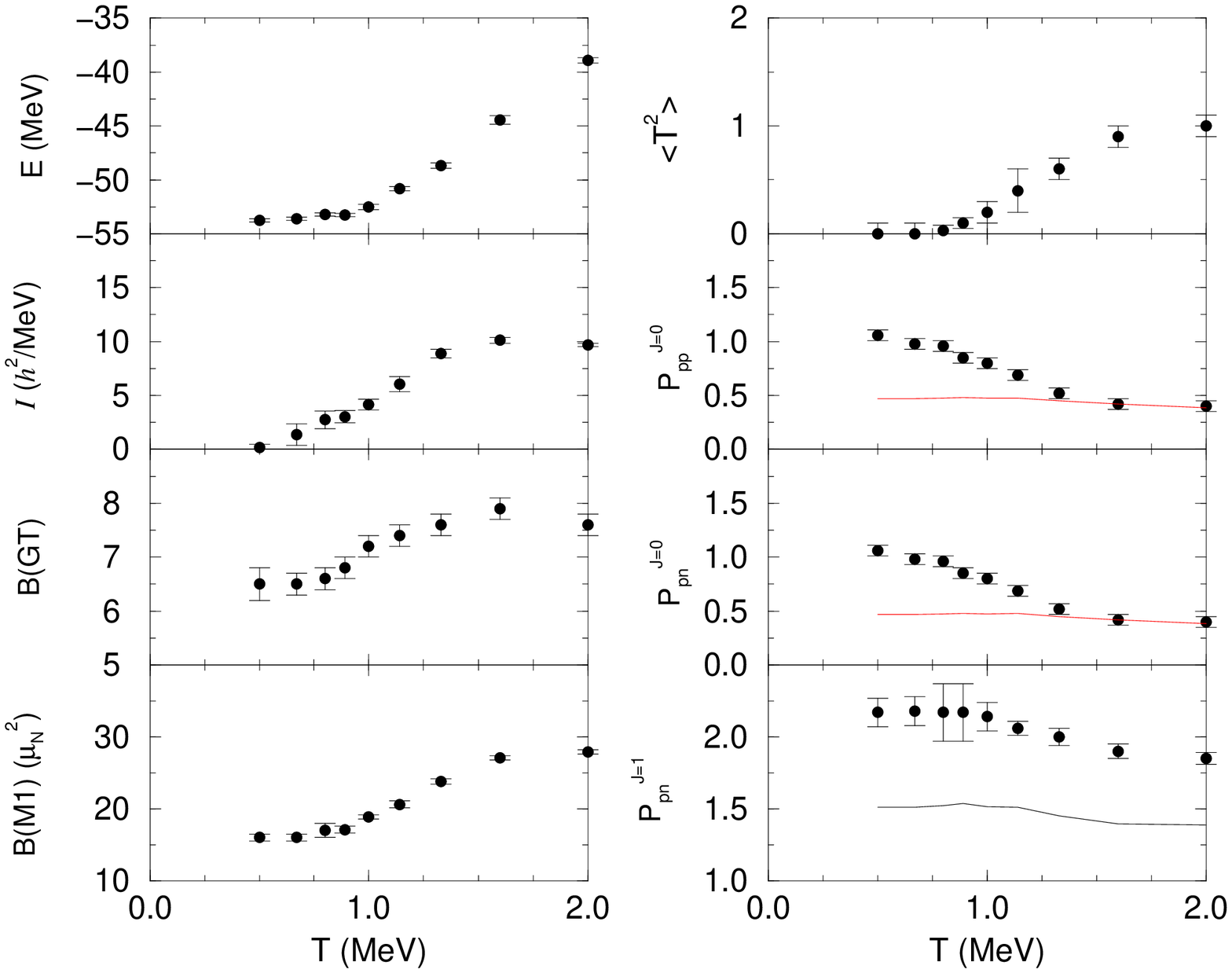}$$ 
\end{center} 
{Fig.~\protect\ref{fe52}. Thermal properties of $^{52}$Fe. The SMMC results
are shown with error bars, while the lines indicate the mean-field values for
the respective pair correlations. } 
\end{figure}

As required by general thermodynamic principles, the internal energy
increases steadily with temperature. The heat capacity $C(T)=dE/dT$, with
$E=\langle H\rangle$, is usually associated with the level density parameter
$a$ by $C(T)=2a(T) T$. As is typical for even-even nuclei \cite{therm} $a(T)$
increases from $a=0$ at T=0 to a roughly constant value at temperatures above
the phase transition. We find $a(T) \approx 5.3\pm1.2$~MeV$^{-1}$ at $T \ge
1$~MeV, in agreement with the empirical value of 6.5~MeV$^{-1}$
\cite{Thielemann} for ${}^{52}$Fe. At higher temperatures, $a(T)$ must
decrease due to the finite model space of our calculation. 

The temperature dependence of $E$ in ${}^{50}$Mn is significantly different
from that in even-even nuclei. As can be seen in Fig.~\ref{mn50}, $E$
increases approximately linearly with temperature, corresponding to a
constant heat capacity $C(T) \approx 5.4\pm1$~MeV$^{-1}$; the level density
parameter decreases like $a(T) \sim T^{-1}$ in the temperature interval
between 0.4~MeV and 1.5~MeV. We note that the same linear increase of the
energy with temperature is observed in SMMC studies of odd-odd $N=Z$ nuclei
performed with a pairing+quadrupole hamiltonian \cite{Zheng}.

To investigate the apparent differences in the thermal behavior of even-even
and odd-odd $N=Z$ nuclei in a more systematic way, we have performed SMMC
calculations for $^{50}$Mn and $^{50}$Cr using a pairing+quadrupole
hamiltonian. Note that the quadrupole part allows for $T=0$ $pn$
correlations. The strength of the quadrupole component has been adjusted to
reasonably reproduce SMMC results for the realistic KB3 interaction. This
hamiltonian has no sign problem, thus significantly reducing the statistical
uncertainties (and also potential systematic errors related to the
g-extrapolation). Despite its simplicity, the hamiltonian nevertheless
embraces much of the essential degrees of freedom governing the thermal
response at low temperatures.

\begin{figure} 
\label{Cv_50} 
\begin{center} 
$$\epsfxsize=4.5truein\epsffile{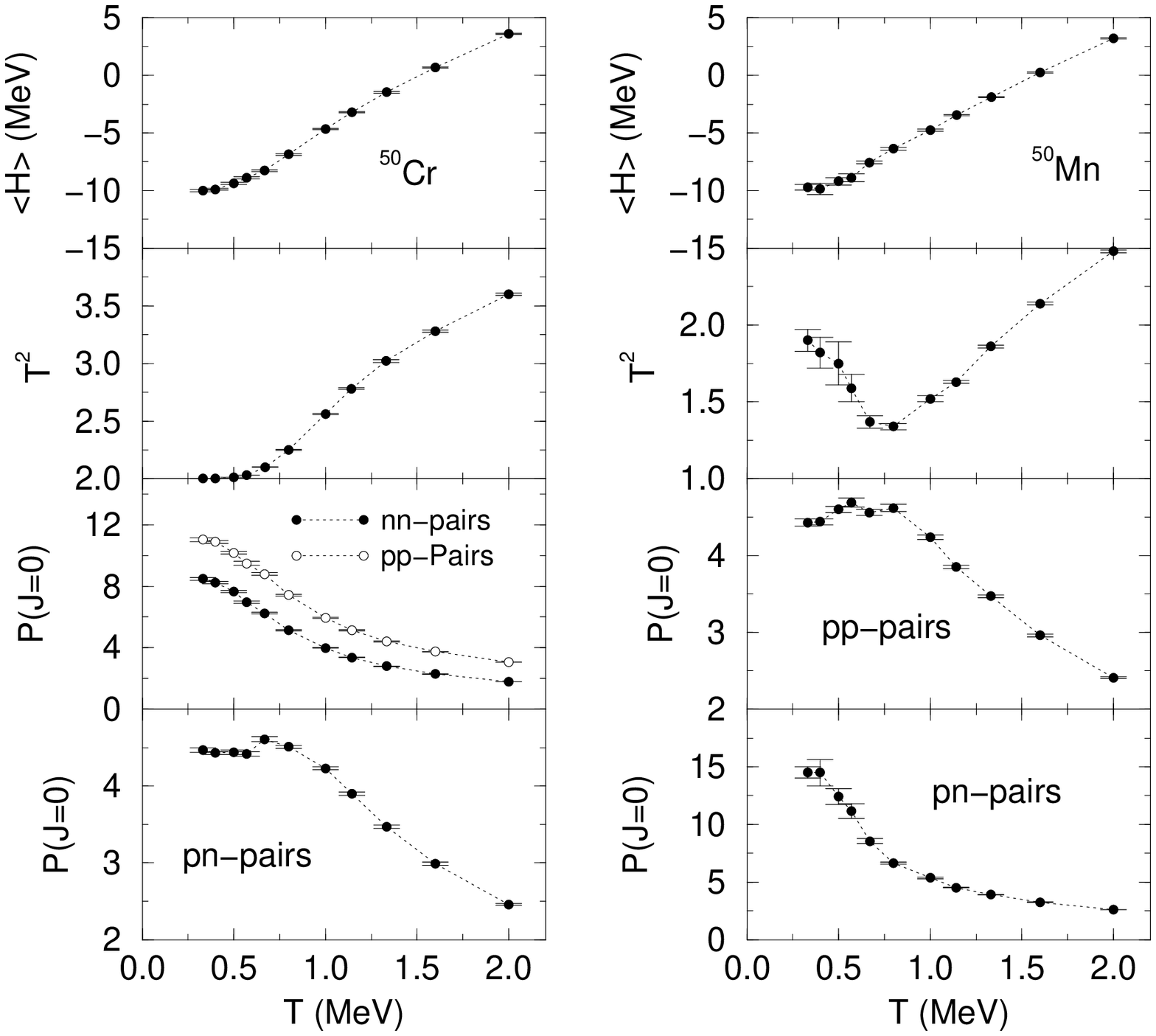}$$ 
\end{center} 
{Fig.~\protect\ref{Cv_50}. Thermal properties of ${}^{50}$Cr (left) and
${}^{50}$Mn (right) calculated with a simple pairing + quadrupole
hamiltonian.} 
\end{figure}

\begin{figure} 
\label{mn50} 
\begin{center} 
$$\epsfxsize=4.5truein\epsffile{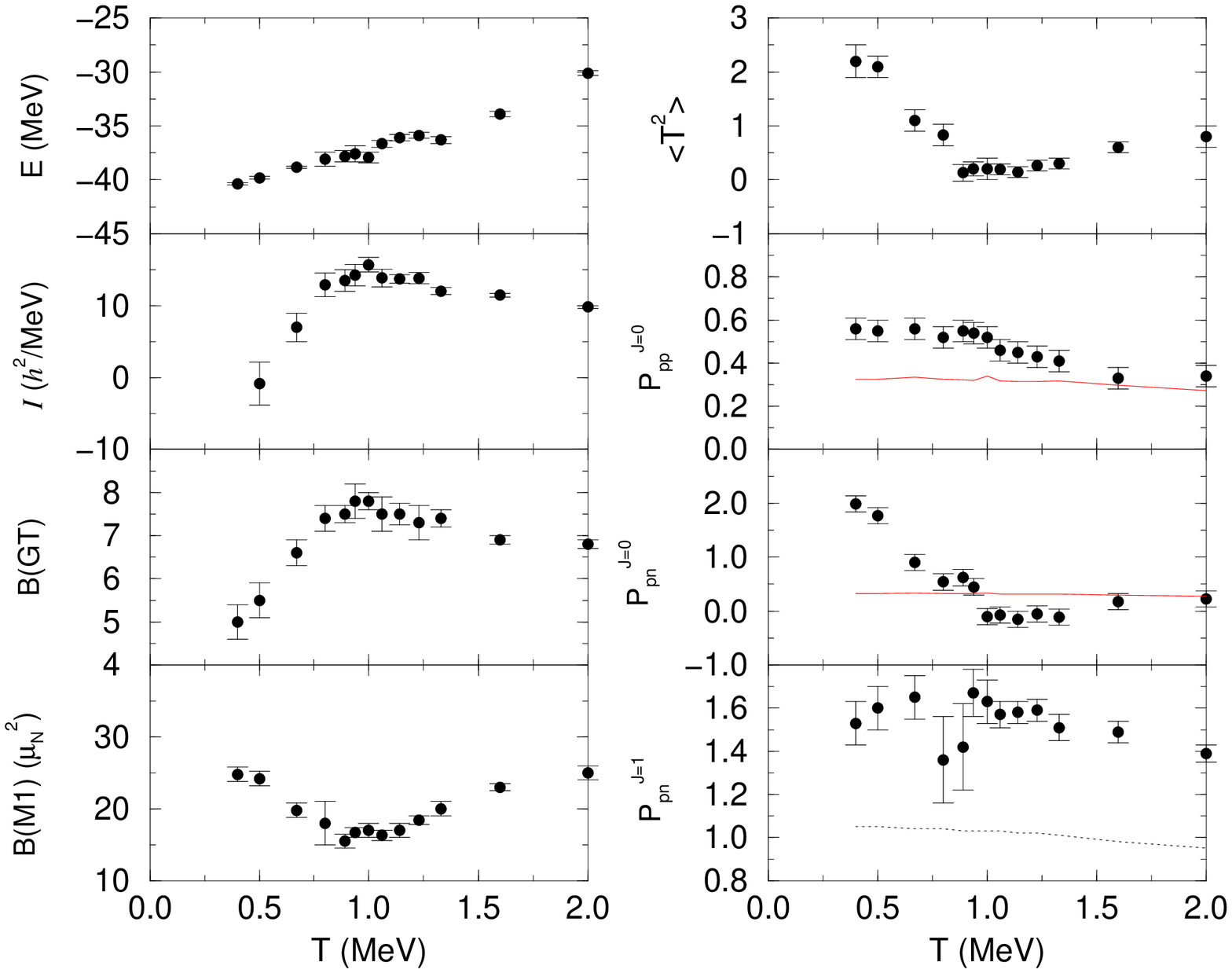}$$ 
\end{center} 
{Fig.~\protect\ref{mn50}. Thermal properties of $^{50}$Mn. The SMMC results
are shown with error bars, while the lines indicate the mean-field values for
the respective pair correlations. } 
\end{figure}

The ground states of ${}^{50}$Cr and ${}^{50}$Mn belong to the same $T=1$
multiplet and are therefore isobaric analogues. This is correctly recovered
in the SMMC calculation as the energy expectation values at low temperatures
are, within error bars, identical. However, $\langle T^2 \rangle =
1.89\pm0.07$ for ${}^{50}$Mn at $T=0.33$~MeV (the lowest temperature at which
we have performed these 
SMMC calculations) indicates some isoscalar components
being mixed in. (Experimentally, 
the lowest $T=0$ state in ${}^{50}$Mn is at an
excitation energy of 0.2~MeV.) With increasing temperature the relative
strength of the isovector component in ${}^{50}$Mn weakens. Correspondingly,
$\langle T^2\rangle$ decreases to $1.34\pm0.04$ at $T=0.8$~MeV. As isospin
$T=0$ states cannot be formed in ${}^{50}$Cr due to the neutron excess
($\langle T^2 \rangle = 2.1$ at $T=0.8$~MeV), this nucleus has fewer degrees
of freedom than ${}^{50}$Mn. As a consequence ${}^{50}$Mn has a higher
excitation energy, or relatedly a higher level density, than ${}^{50}$Cr at
modest temperatures (at $T=0.8$~MeV $\langle H \rangle =-7.55 \pm 0.13$~MeV
for ${}^{50}$Mn and $-8.28\pm0.08$~MeV for ${}^{50}$Cr). At even higher
temperatures, the like-particle correlations break (see Fig.~\ref{Cv_50}).
For both nuclei, isospin states with $T \geq 2$ get mixed in and $\langle
T^2\rangle$ increases. For temperatures above the phase transition ($T\geq
1$~MeV), the thermal properties for both nuclei ($ H , J^2, Q^2, Q_v^2,
Q_p^2, Q_n^2$; the latter two with the appropriate scaling by the proton and
neutron numbers) become almost the same. At $T\leq1$~MeV, the isovector $0^+$
pairing correlations behave as for the realistic KB3 interaction. In
particular, there is the noticeable excess of $pp$ and $nn$ correlations in
the even-even nucleus ${}^{50}$Cr at low temperatures, while the odd-odd
${}^{50}$Mn is dominated by $pn$ correlations. The latter decrease rather
rapidly with temperature, while the $pp$ and $nn$ correlations in ${}^{50}$Cr
show the behavior characteristic of a phase transition near 1~MeV, which is
typical for even-even nuclei in this mass range studied with the KB3
interaction.

However, there are a few differences between calculations with the simple
hamiltonian and the realistic one. First, for the simple hamiltonian all
isovector correlations decrease much more slowly at temperatures exceeding 1
MeV. The origin for this is the missing isoscalar $pn$ correlations, which
dominate nuclear properties at $T \ge 1$ MeV \cite{Langanke}. Their
persistence at these higher temperatures further suppresses the isovector
correlations when the realistic hamiltonian is used. Second, quantities that
are sensitive to $np$ correlations are not correctly described by the simple
hamiltonian. Among these are the Gamow-Teller strengths. Here the realistic
hamiltonian yields $B(GT_+) = 5.2 \pm 1.8$ for $^{50}$Mn and 2.2$\pm$0.2 for
$^{50}$Cr (after the renormalization of the spin operator). In contrast, for
the simple hamiltonian we find $B(GT_+)= 4.5 \pm 0.1$ for $^{50}$Mn and
3.8$\pm$0.01 for $^{50}$Cr. These values have to be compared with the
independent-particle model estimates: 6.9 ($^{50}$Mn) and 5.1 ($^{50}$Cr).
While the large error bars make a comparison for $^{50}$Mn meaningless, the
isovector $pn$ pairing accounts for somewhat less than half of the $GT_+$
quenching in $^{50}$Cr, the reminder being due to isoscalar $pn$ correlations
between spin-orbit partners, introduced by the $(\vec{\sigma} \tau)^2$ piece
missing in the simple hamiltonian. Third, the level density at moderate
temperatures ($T \approx 1$~MeV) is less for the realistic hamiltonian than
for the simple one; one of the reasons again is the missing isoscalar $pn$
correlations that push levels to higher energies.

Nevertheless these SMMC calculations with the sign-problem-free hamiltonian
are quite illustrative and can be extended to odd-$A$ and odd-odd $N \neq Z$
nuclei. First calculations show that one can investigate these nuclei down to
temperatures of order 0.5 MeV, before the sign-problem seriously enters
again. 

\subsection{$\beta\beta$-decay}

The second-order weak process $(Z,A)\rightarrow(Z+2,A)+2e^-+2\bar\nu_e$ is an
important ``background'' to searches for the lepton-number violating
neutrinoless mode, $(Z,A)\rightarrow(Z+2,A)$. The calculation of the nuclear
matrix element for these two processes is a challenging problem in nuclear
structure, and has been done in a full $pf$ model space for only the lightest
of several candidates, ${}^{48}$Ca. P.B.~Radha {\it et al.} have performed
first Monte Carlo calculations of the $2\nu\;\beta\beta$ matrix elements in
very large model spaces \cite{Radha}.

In two-neutrino double $\beta$-decay, the nuclear matrix element of interest
is 
\begin{equation} 
M^{2\nu}\equiv \sum_m {\langle f_0\vert {\bf G}\vert m\rangle \cdot
\langle m\vert {\bf G}\vert i_0\rangle\over E_m-\Omega}\;, \label{neut_a}
\end{equation} 
where $\vert i_0\rangle$ and $\vert f_0\rangle$ are the $0^+$ ground states
of the initial and final even-even nuclei, and $\vert m\rangle$ is a $1^+$
state of the intermediate odd-odd nucleus; the sum is over all such states.
In this expression, ${\bf G}= \hbox{\bf{$\sigma$}}\tau_-$ is the
Gamow-Teller operator for $\beta^-$-decay (i.e., that which changes a neutron
into a proton) and $\Omega=(E_{i_0}+E_{f_0})/2$. A common approximation to
$M^{2\nu}$ is the closure value, 
\begin{equation}
M^{2\nu}={M_c\over {\bar E}} 
\end{equation} 
where $\bar E$ is an average energy dominator and 
\begin{equation} 
M_c\equiv \sum_m \langle f_0\vert {\bf G}\vert m\rangle\langle
m\vert{\bf G}\vert i_0\rangle= \langle f_0\vert{\bf G}\cdot{\bf
G}\vert i_0\rangle\;. 
\end{equation}

SMMC methods can be used to calculate both $M_c$ and $M^{2\nu}$. To do so,
consider the function 
\begin{eqnarray} 
\phi (\tau , \tau') & = & \langle e^{{H}(\tau+\tau')} {\bf
G}^\dagger\cdot{\bf G}^\dagger e^{-{H}\tau} {\bf G}
e^{-{H}\tau'}{\bf G}\rangle \nonumber \\ 
&=&{1\over Z}{\rm Tr}_A\, \left[ e^{-(\beta-\tau-\tau'){H}} {\bf
G}^\dagger\cdot{\bf G}^\dagger e^{-\tau {H}} {\bf G} e^{-\tau'
{H}} {\bf G}\right]\;, 
\end{eqnarray} 
where $Z={\rm Tr}_A\, e^{-\beta {H}}$ is the partition function for the
initial nucleus, $ H$ is the many-body hamiltonian, and the trace is over
all states of the initial nucleus. The quantities
$(\beta-\tau-\tau^{\prime})$ and $\tau$ play the role of the inverse
temperature in the parent and daughter nucleus respectively. A spectral
expansion of $\phi$ shows that large values of these parameters guarantee
cooling to the parent and daughter ground states. In these limits, we note
that $\phi(\tau,\tau^{\prime}=0)$ approaches $e^{-\tau Q} |M_c|^2$, where
$Q=E_i^0-E_f^0$ is the energy release, so that a calculation of
$\phi(\tau,0)$ leads directly to the closure matrix element. If we then
define 
\begin{equation} 
\eta(T,\tau)\equiv \int_0^T d\tau^{\prime} \phi(\tau,\tau^{\prime})
e^{-\tau^{\prime} Q/2}, 
\end{equation} 
and 
\begin{equation} 
M^{2\nu}(T,\tau)\equiv{\eta(T,\tau) M^{*}_{c} \over \phi(\tau,0)}, 
\end{equation} 
it is easy to see that in the limit of large $\tau$,
$(\beta-\tau-\tau^{\prime})$, and $T$, $M^{2\nu}(T,\tau)$ becomes independent
of these parameters and is equal to the matrix element in Eq.~(\ref{neut_a}).

In the first applications, Radha {\it et al.} calculated the $2\nu$ matrix
elements for ${}^{48}$Ca and ${}^{76}$Ge \cite{Radha}. The first nucleus
allowed a benchmarking of the SMMC method against direct diagonalization. A
large-basis shell model calculation for ${}^{76}$Ge has long been waited for,
as $^{76}$Ge is one of the few nuclei where the $2\nu\beta\beta$ decay has
been measured precisely and the best limits on the $0\nu$ decay mode have
been established \cite{Ge1,Ge2,Klapdor}.

To monitor the possible uncertainty related to the $g$-extrapolation in the
calculation of the $2\nu$ matrix element for ${}^{76}$Ge, SMMC studies have
been performed for two quite different families of sign-problem-free
hamiltonians ($\chi=\infty$ and $\chi=4$). The calculation comprises the
complete $(0f_{5/2},1p,0g_{9/2})$ model space, which is significantly larger
than in previous shell model studies \cite{Haxton}. The adopted effective
interaction is based on the Paris potential and has been constructed for this
model space using the $Q$-box method developed by Kuo \cite{Kuo95}.

\begin{figure} 
\label{bb_decay} 
$$\epsfxsize=4truein\epsffile{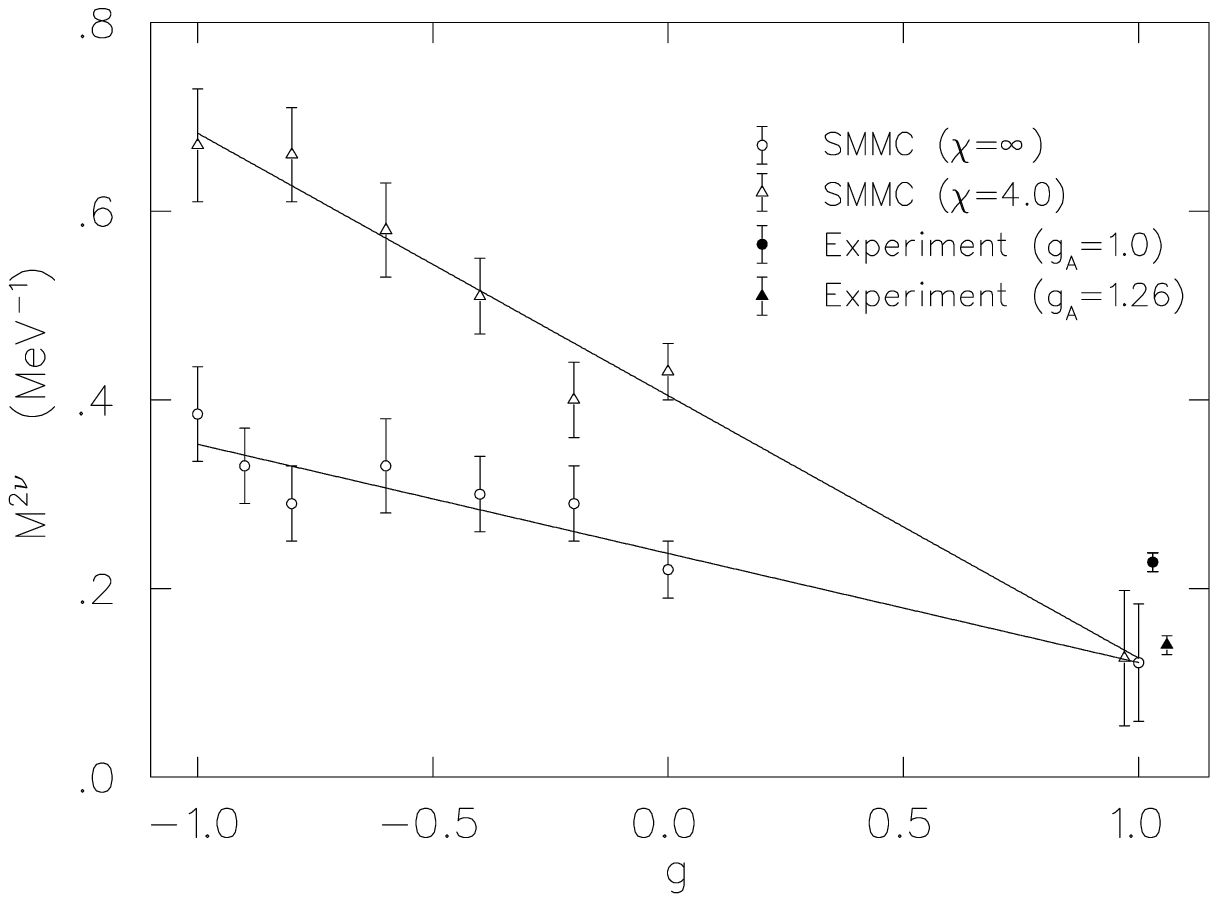}$$ 
{FIG~\protect\ref{bb_decay}. The $2\nu$ matrix element for $^{76}$Ge
calculated within SMMC studies based on two families of hamiltonians which
are free of sign problems. The physical values are obtained by linear
extrapolation to $g=1$. The experimental value for this matrix element
\protect\cite{Klapdor} is indicated by the diamond (from
\protect\cite{Radha}). } 
\end{figure}

As is shown in Fig.~\ref{bb_decay}, upon linear extrapolation, both families
of hamiltonians predict a consistent value for the $2\nu$ matrix element of
${}^{76}$Ge. The results $M^{2\nu}=0.12\pm0.07$ and $M^{2\nu}=0.12\pm0.06$
are only slightly lower than the experimental values ($M^{2\nu}=0.22\pm0.01$
\cite{Klapdor}). This comparison, however, should not be overinterpreted, as
the detailed reliability of the effective interaction is still to be checked.

It is interesting that the closure matrix element found in the SMMC
calculation and the average energy denominator ($M_c=-0.36\pm0.37$, ${\bar
E}=-3.0\pm 3.3$~MeV and $M_c=0.08\pm0.17$, ${\bar E}=0.57\pm 1.26$~MeV for
the two families of hamiltonians with ${\chi}=\infty$ and ${\chi}=4$,
respectively) are both significantly smaller than had been assumed
previously. This is confirmed by a recent truncated diagonalization study
\cite{Caurier96}.

\subsection{$\gamma$-soft nuclei}

Nuclei with mass number $100\leq A \leq140$ are believed to have large shape
fluctuations in their ground states. Associated with this softness are
spectra with an approximate $O(5)$ symmetry and bands with energy spacings
intermediate between rotational and vibrational. In the geometrical model
these nuclei are described by potential energy surfaces with a minimum at
$\beta \neq 0$ but independent of $\gamma$ \cite{8.7F}. Some of these nuclei
have been described in terms of a quartic five-dimensional oscillator
\cite{8.7G}. In the Interacting Boson Model (IBM) they are described by an
$O(6)$ dynamical symmetry \cite{Arima,8.7I,8.7J}. In the following we review
the first fully microscopic calculations for soft nuclei with $100\leq A
\leq140$ \cite{8.7K}.

For the two-body interaction we used a monopole ($J=0)$ plus quadrupole
$(J=2)$ force \cite{8.7L} supplemented by a collective quadrupole
interaction: 
\begin{equation} 
 H_2 = - \sum_{\lambda \mu}{{\pi g_\lambda}\over{2\lambda+1}} 
P^\dagger_{\lambda \mu}  P_{\lambda \mu} -{1\over 2} \chi : \sum_\mu
(-)^{\mu}  Q_\mu  Q_{-\mu}: \;, 
\end{equation} 
where $::$ denotes normal ordering. The single particle energies and the
other parameters were determined as described in Ref.~\cite{8.7K}.

We begin discussion of our results with the probability distribution of the
quadrupole moment. This is obtained from the shape of each Monte Carlo
sample, including the quantum-mechanical fluctuations through the variance of
the ${ Q}$ operator for each sample, $\Delta_{\sigma}^2 = {\rm Tr} ({
Q}^2 { U}_\sigma)/ {\rm Tr} ({ U}_\sigma) -\langle {
Q}\rangle_\sigma^2 $. The shape distribution $P(\beta,\gamma)$ can be
converted to a free energy surface as discussed in Ref.~\cite{8.7K}.

\begin{figure} 
\label{shapes} 
$$\epsfxsize=4truein\epsffile{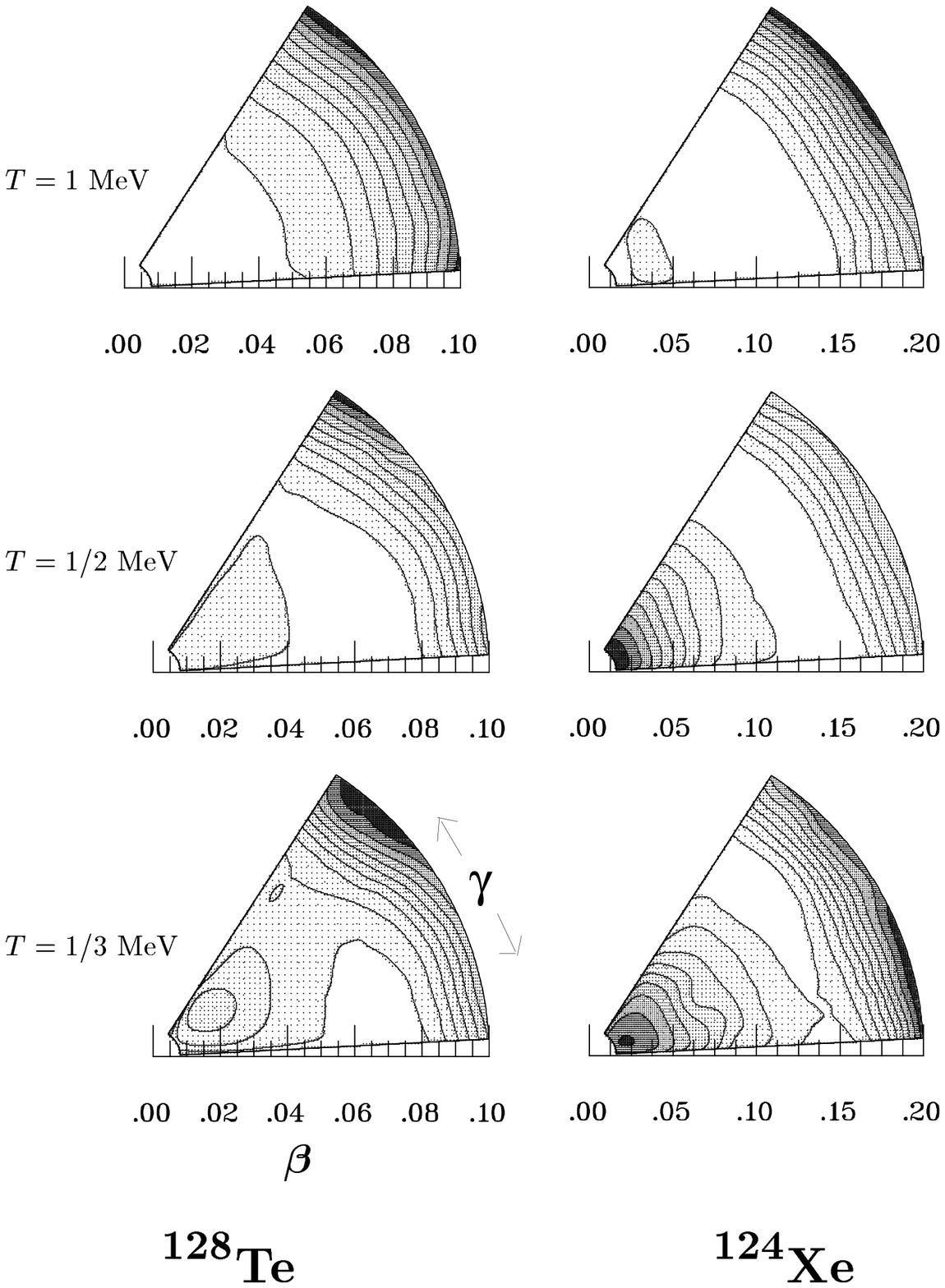}$$ 
{FIG~\protect\ref{shapes}. Contours of the free energy (as described in the
text) in the polar-coordinate $\beta-\gamma$ plane for ${}^{128}$Te and
${}^{124}$Xe. Contours are shown at 0.3 MeV intervals, with lighter shades
indicating the more probable nuclear shapes (from \protect\cite{8.7K}). } 
\end{figure}

The shape distributions of ${}^{128}$Te and ${}^{124}$Xe are shown in
Fig.~\ref{shapes} at different temperatures. These nuclei are clearly
$\gamma$-soft, with energy minima at $\beta \sim 0.06$ and $\beta \sim 0.15$,
respectively. Energy surfaces calculated with Strutinsky-BCS using a deformed
Woods-Saxon potential \cite{8.7N} also indicate $\gamma$-softness with values
of $\beta$ comparable to the SMMC values. These calculations predict for
${}^{124}$Xe a prolate minimum with $\beta \approx 0.20$ which is lower than
the spherical configuration by 1.7 MeV but is only 0.3 MeV below the oblate
saddle point, and for ${}^{128}$Te a shallow oblate minimum with $\beta
\approx 0.03$. These $\gamma$-soft surfaces are similar to those obtained in
the $O(6)$ symmetry of the IBM, or more generally when the hamiltonian has
mixed $U(5)$ and $O(6)$ symmetries but a common $O(5)$ symmetry. In the Bohr
hamiltonian, an $O(5)$ symmetry occurs when the collective potential energy
depends only on $\beta$ \cite{8.7F}. The same results are consistent with a
potential energy $V(\beta)$ that has a quartic anharmonicity \cite{8.7G}, but
with a negative quadratic term that leads to a minimum at finite $\beta$.

The total E2 strengths were estimated from $\langle  Q^2\rangle$ where
${ Q}=e_p  Q_p + e_n  Q_n$ is the electric quadrupole operator
with effective charges of $e_p=1.5 e$ and $e_n=0.5 e$, and $B(E2;
0^+_1\rightarrow 2^+_1)$ determined by assuming that most of the strength is
in the $2_1^+$ state. Values for $B(E2; 0^+_1\rightarrow 2^+_1)$ of $663\pm
10$, $2106\pm 15$, and $5491\pm 36$ $e^2 {\rm fm}^4$ were found, to be
compared with the experimental values \cite{8.7O} of 1164, 3910, and 9103
$e^2 {\rm fm}^4$ for ${}^{124}$Sn, ${}^{128}$Te, and ${}^{124}$Xe,
respectively. Thus, the SMMC calculations reproduce the correct qualitative
trend. The $2_1^+$ excitation energies were also calculated from the E2
response function. The values of $1.12 \pm 0.02$, $0.96 \pm 0.02$, and $0.52
\pm 0.01$ MeV are in close agreement with the experimental values of 1.2, 0.8,
and 0.6 MeV for ${}^{124}$Sn, ${}^{128}$Te, and ${}^{124}$Xe, respectively.

\begin{figure} 
\label{gam_crank} 
$$\epsfxsize=4truein\epsffile{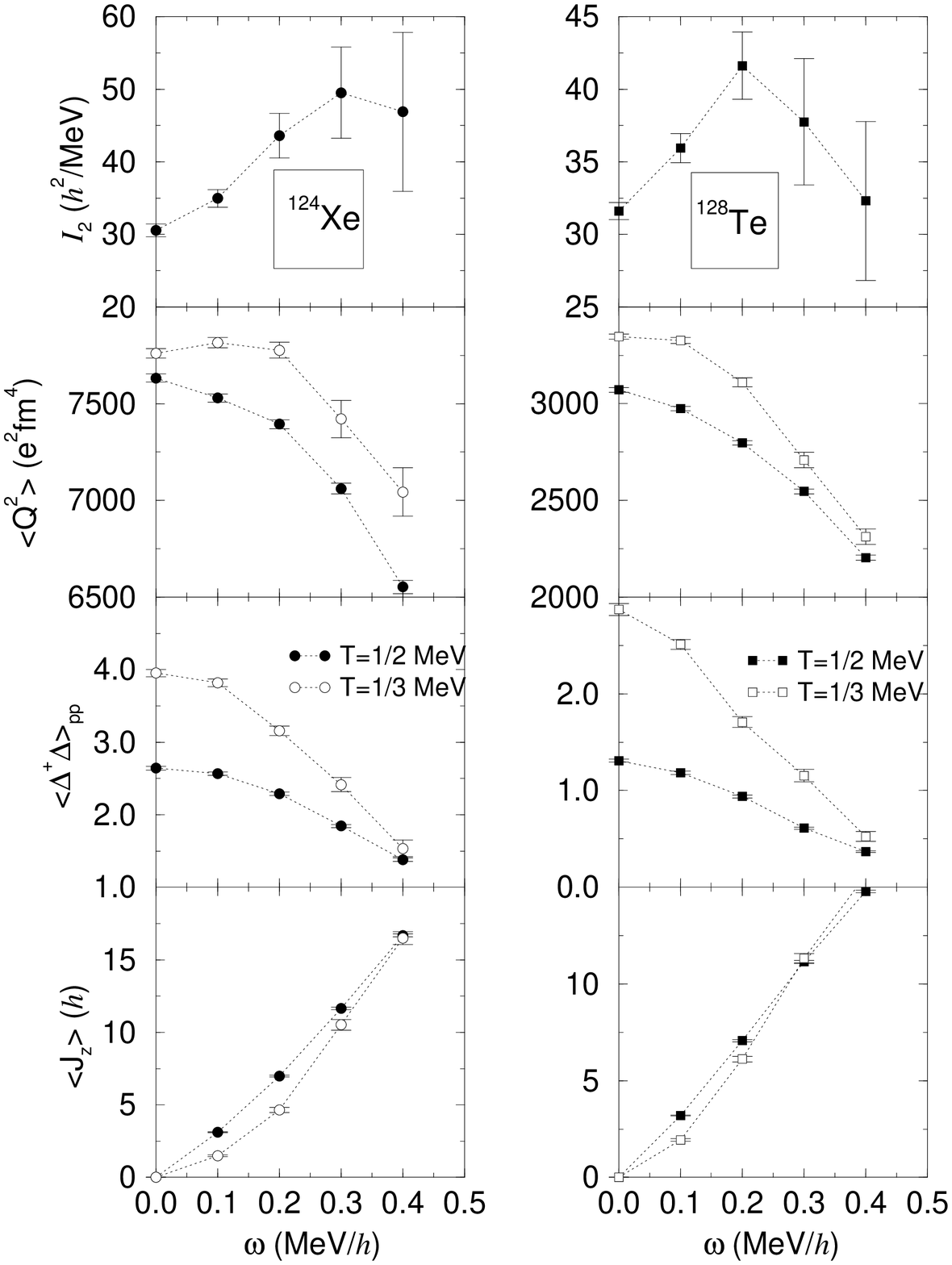}$$ 
{FIG~\protect\ref{gam_crank} Observables for ${}^{124}$Xe and ${}^{128}$Te as
a function of cranking frequency $\omega$ and for two temperatures. ${\cal
I}_2$ is the moment of inertia, $Q$ is the mass quadrupole moment, $\Delta$
is the $J=0$ pairing operator, and $J_z$ is the angular momentum along the
cranking axis (from \protect\cite{8.7K}). } 
\end{figure}

Another indication of softness is the response of the nucleus to rotations,
probed by adding a cranking field $\omega J_z$ to the hamiltonian and
examining the moment of inertia as a function of the cranking frequency
$\omega$. For a soft nucleus one expects a behavior intermediate between a
deformed nucleus, where the inertia is independent of the cranking frequency,
and the harmonic oscillator, where the inertia becomes singular. This is
confirmed in Fig.~\ref{gam_crank} which shows the moment of inertia ${\cal
I}_2$ for ${}^{124}$Xe and ${}^{128}$Te as a function of $\omega$, and
indicates that $^{128}$Te has a more harmonic character. The moment of
inertia for $\omega=0$ in both nuclei is significantly lower than the rigid
body value ($\approx 43 \hbar^2/$MeV for $A=124$) due to pairing
correlations.

Also shown in Fig.~\ref{gam_crank} are $\langle  Q^2 \rangle$ where $
Q$ is the mass quadrupole, the BCS-like pairing correlation $\langle 
\Delta^\dagger  \Delta\rangle$ for the protons and $\langle 
J_z\rangle$ (neutron pairing is less affected, and therefore not shown).
Notice that the increase in $I_2$ as a function of $\omega$ is strongly
correlated with the rapid decrease of pairing correlations and that the peaks
in $I_2$ are associated with the onset of a decrease in collectivity (as
measured by $\langle  Q^2 \rangle$). This suggests band crossing along
the yrast line associated with pair breaking and alignment of the
quasi-particle spins at $\omega \approx 0.2$ MeV ($\langle  J_z \rangle
\approx 7 \hbar$) for ${}^{128}$Te and $\omega \approx 0.3$ MeV ($\langle
 J_z \rangle \approx 11 \hbar$) for ${}^{124}$Xe. The results are
consistent with an experimental evidence of band crossing in the yrast
sequence of ${}^{124}$Xe around spin of 10 $\hbar$ \cite{8.7P}. The alignment
effect is clearly seen in the behavior of $\langle  J_z \rangle$ at the
lower temperature, which shows a rapid increase after an initial moderate
change. Deformation and pairing decrease also as a function of temperature.

The total number of $J$-pairs ($n_J=\sum_\alpha n_{\alpha J}$) in the various
pairing channels was also calculated. Since the number of neutrons in
${}^{124}$Xe is larger than the mid-shell value, they are treated as holes.
For $J=0$ and $J=2$, one can compare the largest $n_{\alpha J}$ with the
number of $s$ and $d$ bosons obtained from the $O(6)$ limit of the IBM. For
${}^{124}$Xe the SMMC (IBM) results in the proton-proton pairing channel are
0.85 (1.22) $s$ ($J=0$) pairs, and 0.76 (0.78) $d$ ($J=2$) pairs, while in
the neutron-neutron channel we find 1.76 (3.67) $s$ pairs and 2.14 (2.33) $d$
pairs. For the protons the SMMC $d$ to $s$ pair ratio 0.89 is close to its
$O(6)$ value of 0.64. However, the same ratio for the neutrons, 1.21, is
intermediate between $O(6)$ and $SU(3)$ (where its value is 1.64) and is
consistent with the neutrons filling the middle of the shell. The total
numbers of $s$ and $d$ pairs---1.61 proton pairs and 3.8 neutron (hole)
pairs---are below the IBM values of 2 and 6, respectively.

\section{Illustrative calculations}

\subsection{Giant dipole resonances}

To describe the GDR in ${}^{16}$O, the $0p$-$1s0d$-$0f_{7/2}$ shell model
space was used \cite{ormand94}, and the residual interaction was composed of
pairing and multipole terms with the specific parametrization given in
\cite{ormand94},
\begin{equation} 
V_{\rm res} =\chi_1 (q^1_p -q^1_n)^2 +\chi_2 (q^2_p + q^2_n)^2 + \chi_4
(q^4_p +q^4_n)^2\;. 
\end{equation} 

\begin{figure} 
\label{fig_dipole} 
$$\epsfxsize=4truein\epsffile{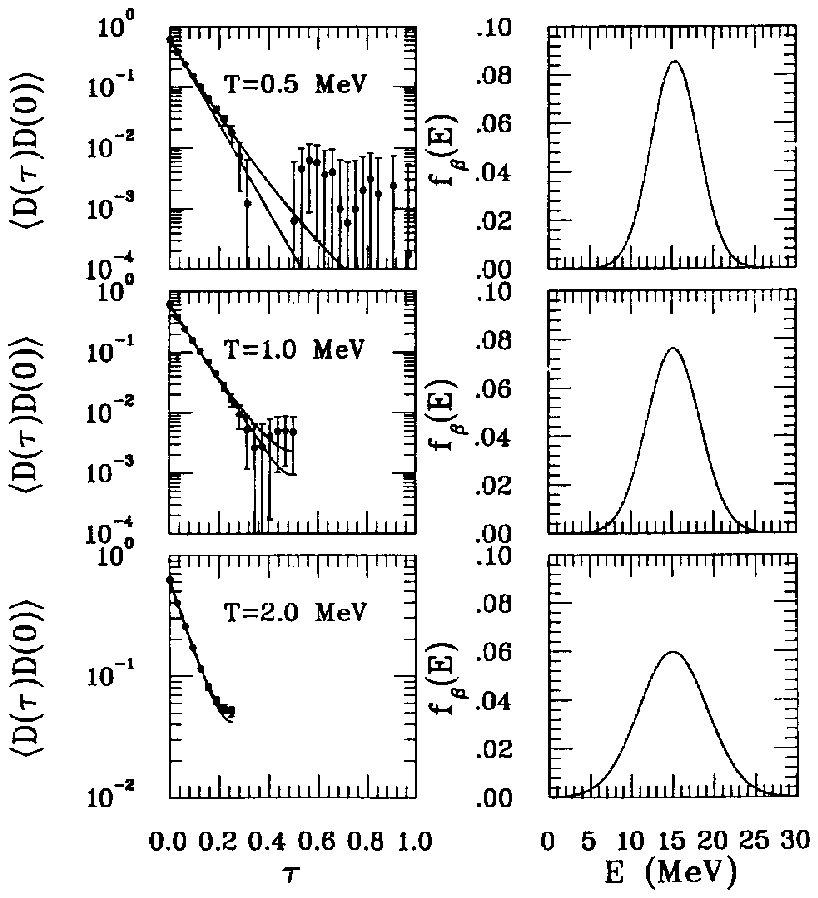}$$ 
{FIG~\protect\ref{fig_dipole} Response function (left panels)
and their excitation strength distributions (right panels)
for isovector dipole transitions in $^{16}$O 
at $T=0.5$, $1.0$, and $2.0$~MeV. }
\end{figure}

Shown in Fig.~\ref{fig_dipole} are the results obtained for $^{16}$O for the
dipole transitions $D=1/2(q^1_p - q^1_n)$ at temperatures $T=$0.5, 1.0, and
2.0~MeV. Although these calculations require further refinement, some general
remarks are warranted. First, from the values of $R(\tau)$ at $\tau=0$, the
total dipole strength is roughly constant as a function of temperature, as is
the first moment of the strength function. The width, or second moment, cannot
be reliably extracted from the calculations presented in
Fig.~\ref{fig_dipole} due to large Monte Carlo errors. Still, the
calculations are representative of what may be obtained with more samples
(yielding smaller errors). A future study will involve a more complete model
space description, and implementation of matrix-multiply stabilization so
that lower temperatures may be studied.

\subsection{Multi-major shell calculations}

Numerous problems in nuclear physics would benefit from the ability to
calculate shell model observables in more than a single oscillator shell. For
example, the structure of light neutron-rich nuclei in the $sd$-shell requires
inclusion of $fp$-shell orbitals in order to properly describe their
deformation and cross shell characteristics. Experimentally, this region will
be probed to understand the nature of weakly bound systems, and the response
of nuclei near the neutron drip line to various physical probes.
Multi-$\hbar\omega$ SMMC calculations should be no more difficult
computationally than the $0\hbar\omega$ calculations that are now routine,
although they do require increased memory and computational cycles.

When more than one oscillator shell is included, a new complication arises:
excitations of the center-of-mass (CM) become mixed into the real nuclear
excitations. This leads to spurious components of wave functions being
incorporated into the calculations of observables. We have investigated two
methods of removing these spurious components in the SMMC. Projecting out
these components leads to another sign problem and has proven unworkable. The
Gloeckner-Lawson prescription \cite{GL} has been shown to work for fairly
small CM multipliers $(< 5)$ in comparisons with direct diagonalization
calculations. Unfortunately, this method requires a ``complete''
$n\hbar\omega$ space to be fully effective and SMMC methods cannot work in
such a complete space \cite{MW75}. The Gloeckner-Lawson prescription will
still remove the spurious solutions at the cost of also slightly affecting
the non-spurious solutions. The best value of the multiplier is currently
being investigated. Values of order 1 seem to be best.

\begin{figure} 
\label{center_of_mass} 
\epsfxsize=4.0truein\epsffile{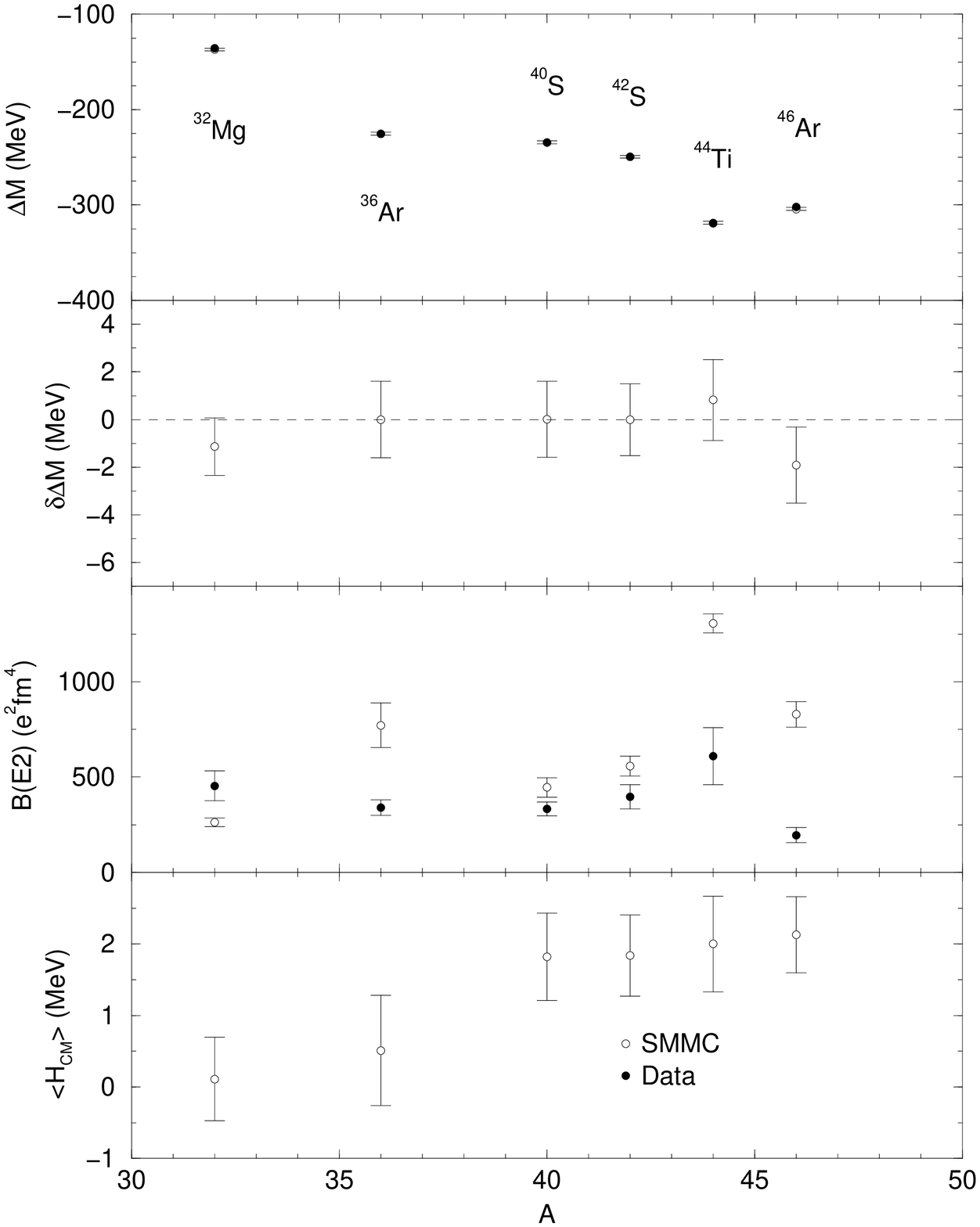} 
\vskip0.2in
{FIG~\protect\ref{center_of_mass} A selection of SMMC calculated observables
vs experiment. A) The calculated and experimental mass excess. B) Deviation
between experimental and calculated masses. C) The total calculated $B(E2)$
vs. the experimentally measured $B(E2)$ to the first excited 2$^+$ state. D)
The expectation value of the center of mass hamiltonian. The values are much
smaller than $2\hbar\omega$, the minimum value for a spurious wave function.
} 
\end{figure}

Our current research focuses on the $N = 20$ and 28 shell closures for very
neutron-rich nuclei where we may investigate how exact the shell closure
really is for nuclei such as ${}^{32}$Mg and ${}^{46}$Ar. 
(In the $sdfp$ shell
this calculation allows excitations of up to $16\hbar\omega$ for
${}^{32}$Mg.) Experimentally, ${}^{32}$Mg is highly deformed \cite{mg32}. An
$sd$-shell calculation gives roughly 30\% of this deformation, while the
inclusion of the $pf$-shell yields much of the remaining deformation due to
two-particle excitations into the $pf$-shell. There is no well-established
interaction for the $sdpf$ shell, so we have chosen to use the WBMB
interaction of Ref.~\cite{wbb}. In order to find agreement with the
experimental mass excess the single particle energies of the $pf$ shell were
lowered by 4~MeV relative to the $sd$ shell. We are interested in both the
deformation of ${}^{32}$Mg, and the neutron-rich sulfur isotopes which have
recently been experimentally measured\cite{scheit96}. 
In Fig.~\ref{center_of_mass} we show for several nuclei in the region, 
the calculated
mass excess for this interaction compared to experiment. We also show the
total calculated $B(E2)$ vs. the experimentally measured $B(E2)$ to the first
$2^{+}$ state. The final panel shows the expectation value of the 
center-of-mass 
hamiltonian. For all of the nuclei, a spurious $2\hbar\omega$ solution
would have $\langle H_{\rm CM} \rangle \approx 20$~MeV. The center-of-mass
contamination is seen to be quite small.

\section{Concluding remarks}

We have demonstrated in the preceding two sections the powers and
limitations of the SMMC technique in describing nuclear properties. The SMMC
technique, while often non-trivial to apply to a given situation, indeed sheds
light on various aspects of nuclear systems. The most important studies
mentioned above include the quenching of the Gamow-Teller strength, thermal
and rotational properties of pair correlations, and the evolution of nuclear
shapes with temperature.

We have witnessed enormous computational advances in the last several years.
Large-scale parallel computing for scientific problems continues to move
forward quickly. Thus some of the more difficult problems that have not been
treated by SMMC at present could be pursued in the near future. Among these
is the GDR in a several $\hbar\omega$ space; studies of nuclei in the $Pb$
region; studies of weakly bound systems; and studies at very low temperatures
[requiring matrix-multiply stabilization discussed in Ref.~\cite{Koonin96}].
The low temperature studies would allow for a detailed description of
strength functions not currently available to us. The advent of radioactive
beam experiments will also allow for interesting new physics to be studied
within the realm of the shell model.

The ground state and thermal properties of nuclear matter are another
intriguing application of SMMC methods. One approach is to use single
particle states that are plane waves with periodic boundary conditions, and a
$G$-matrix derived from a realistic inter-nucleon interaction; the formalism
and algorithms we have presented here are then directly applicable. An
alternative approach is to work on a regular lattice of sites in coordinate
space and employ Skyrme-like effective interactions that couple neighboring
sites; the calculation is then similar to that for the Hubbard model for
which special techniques must be used to handle the large, sparse matrices
involved \cite{Linden}. Both approaches are currently being pursued.

Otsuka and collaborators \cite{Otsuka} have recently proposed a hybrid scheme
whereby some of the SMMC sampling methods are used to select a many-body
basis, which is then employed in a conventional diagonalization. The brunt of
their effort involves finding the local Hartree-Fock minima of the system,
and then sampling around those minima. $J$-projection is approximately
incorporated. With this approach one is able to obtain the lowest few
discrete levels and transitions. The method is a way of truncating the
many-body basis before diagonalization, and is thus complementary to SMMC in
the same way as the conventional shell model.

The work discussed above is the culmination of the efforts of many people
including Y.~Alhassid, C.~Johnson, G.~Lang, E.~Ormand, P.B.~Radha, P.~Vogel,
and more recently M.T.~Ressell, and J.~White. This work was supported in part
by the National Science Foundation, Grant Nos. PHY94-12818 and PHY94-20470.
Oak Ridge National Laboratory (ORNL) is managed by Lockheed Martin Energy
Research Corp. for the U.S. Department of Energy under contract number
DE-AC05-96OR22464. DJD acknowledges an E.P. Wigner Fellowship from ORNL.
Compuational resources were provided by the Center for Advanced Computational
Research at Caltech, the Maui High Performance Computing Center, The RIKEN
computer center, and Center for Computational Sciences at ORNL.

\end{document}